\documentclass[12pt,epsf]{article}
\usepackage{hyperref}
\usepackage{subfigure}
\usepackage{graphics}
\usepackage{amssymb}
\usepackage{epsfig}
\usepackage{rotating}
\usepackage{pstricks}

\def\l{\lambda}

\def\g{\gamma}

\def\ds#1{#1\kern-1ex\hbox{/}}
\def\dsh{h\kern-1.2ex /}

  \let\D=\Delta

\newcommand{\bea}{\begin{eqnarray}}
\newcommand{\eea}{\end{eqnarray}}

\def\nn{\nonumber}
\def\beq{\begin{equation}}
\def\eeq{\end{equation}}

\def\beqn{\begin{eqnarray}}
\def\eeqn{\end{eqnarray}}
\def\ba{\begin{eqnarray}}
\def\ea{\end{eqnarray}}

\def\slash#1{#1\hskip-6pt/\hskip6pt}

\newcommand{\be}{\begin{equation}}
\newcommand{\beqa}{\begin{eqnarray}}
\newcommand{\eeqa}{\end{eqnarray}}
\newcommand{\eps}{\epsilon}
\newcommand{\la}{\lambda}
\newcommand{\ro}{\rho}
\newcommand{\si}{\sigma}
\newcommand{\veps}{\varepsilon}
\newcommand{\ee}{\end{equation}}

\begin{document}
\begin{center}
\vspace{05.cm}
{\bf  \large  Anomalous $U(1)$ Models in Four and Five Dimensions \\ and their Anomaly Poles}

\vspace{1.0cm}
{\bf Roberta Armillis, Claudio Corian\`{o}, Luigi Delle Rose and
 Marco Guzzi\footnote{roberta.armillis@le.infn.it,claudio.coriano@le.infn.it,luigi.dellerose@le.infn.it,marco.guzzi@le.infn.it}}

\vspace{0.5cm}

{\it  Dipartimento di Fisica, Universit\`{a} del Salento \\
and  INFN Sezione di Lecce,  Via Arnesano 73100 Lecce, Italy}\\
\vspace{.5cm}

\begin{abstract}
We analyze the role played by anomaly poles in an anomalous gauge theory by discussing their signature in the corresponding off-shell effective action. The origin of these contributions, in the most general kinematical case, is elucidated by performing a complete analysis of the anomaly vertex at perturbative level. We use two independent (but equivalent) representations: the Rosenberg representation and the longitudinal/transverse (L/T) parameterization, used in recent studies of $g-2$ of the muon and in the proof of non-renormalization theorems of the anomaly vertex. The poles extracted from the L/T parameterization do not couple in the infrared for generic anomalous vertices, as in Rosenberg, but we show that they are responsible for the violations of
unitarity in the UV region, using a class of pole-dominated amplitudes. We conclude that consistent formulations of anomalous models require necessarily the cancellation of these polar contributions. Establishing the UV significance of these terms provides a natural bridge between the anomalous effective action and its completion by a nonlocal theory.  Some additional difficulties with unitarity of the mechanism of inflow in extra dimensional models with an anomalous theory on the brane, due to the presence of anomaly poles, are also pointed out.

 \end{abstract}
\end{center}
\newpage
\section{Introduction and Summary}
One of the subtle features of the axial anomaly is the presence of massless poles in the corresponding AVV correlator, which show up in special kinematical regions and in the chiral limit, and whose interpretation is at times rather puzzling. In fact, on several occasions the correct interpretation of these singularities have been debated at length \cite{Achasov:1992hg,Horejsi:1994aj}. Our interest in the topic, which is one of our reasons and motivations for this analysis, has been the result of a recent work in which we have suggested the subtraction of the anomaly pole in theories involving anomalous $U(1)$'s  to ensure anomaly cancellation, by defining a new gauge invariant vertex \cite{Armillis:2008bg}. The re-defined vertex is non-local, while its Ward identity is expressed in terms of local interactions and can
be interpreted diagrammatically by introducing a massless pseudoscalar - an axion field - coupled to gauge fields via Wess-Zumino terms. This coupling is induced by the anomaly and the subtraction of the anomaly pole is expected to represent the only consistent way by which a  completion of an anomalous theory is supposed to work in the UV region.
 
However, as known from several previous studies of this
vertex, the presence of a longitudinal pole in an anomaly diagram has always been established
{\em only for special kinematical configurations} and this raises a serious concern regarding the meaning of the subtraction, introduced to restore the Ward identity at high energy, a subtraction which should be 
naturally performed by the UV completion of the anomalous theory. The main objective of this work is to show 
that the effective action of an anomalous gauge theory is affected by singularities which are not necessarily detected using a dispersive analysis in the infrared (IR) \cite{Coleman:1982yg} (see also 
\cite{Giannotti:2008cv} for a recent study), and as such are IR decoupled.
These additional poles, which account for the anomaly, can be extracted by a complete computation of the effective action and have a direct ultraviolet UV significance. For this reason, assessing the UV significance 
of an anomaly pole, whose identification, in the past, has always been linked to the infrared (IR) 
using a spectral approach, certainly helps in establishing a natural link between an anomalous theory and its completion, which should guarantee the cancellation of these contributions. 

To show the existence of these singularities under the most general kinematical conditions we proceed with a complete and comparative 
 study of the anomaly diagram in two different parameterizations which are both essential in order to understand the nature of the longitudinal subtraction. In fact, only a complete and off-shell computation of the effective action for an anomalous theory allows the identification of these terms which escape detection with the 
 usual spectral analysis. The nature of these additional singularities of the effective action which, in some cases, are not evident due to the presence of Schouten relations, is resolved by studying a special class of amplitudes in which the presence of a pole dominance can be immediately linked to a non unitary behaviour of the theory.  
Having clarified these points, we proceed by discussing the structure of the anomalous effective action of a typical  anomalous theory, represented by expansions in the fermion mass ($m$). This can be viewed as the generalization to the anomalous case of the usual Euler-Heisenberg effective action, which now contains additional (anomalous) trilinear interactions that are absent in the QED case, due to C-invariance.

Then we turn to a brief discussion of anomaly poles in theories with extra dimensions. In this case we briefly point out that a mechanism of inflow which does not erase the anomaly poles of the effective (anomalous) theory localized on the brane, may run into additional difficulties with unitarity, besides the well-known ones \cite{SekharChivukula:2001hz,Chivukula:2003kq, Muck:2001yv} which imply a truncation on the KK modes. The example of a simple $S_1/Z_2$ compactification of a 5-D gauge theory with an inflow generated by a 5-D Chern-Simons term,
following closely the construction of \cite{Hill:2006ei}, is brought up to illustrate our point. 
We conclude with some perspectives on how to extend our studies of pole dominance to other situations, such as in the conformal anomaly, which could help in supporting quite independently our results.

\section{Anomaly poles and general kinematics: the Rosenberg case}
One of the intriguing features of the anomaly diagrams is that the poles are part of the anomaly
amplitude only under some special kinematical conditions. For instance, the $\pi\to \gamma \gamma$ (pion pole) amplitude interpolates between the axial vector current ($J_A$) and two vector currents ($J_V$) and saturates the anomaly contribution (if we neglect the pion mass)
given by the $\langle J_A J_V J_V \rangle$ perturbative correlator. This saturation is at the basis of 't Hooft's  matching conditions, according to which the anomaly of the fermions should be reproduced by a composite particle (a pseudoscalar) in a confining theory (see also the discussion in  \cite{Giannotti:2008cv}).  In general, the pole appears by solving the anomalous Ward identity for the corresponding amplitude,  $\Delta^{\lambda\mu\nu} (k_1,k_2) $ (we use momenta as in Fig.~\ref{AVV}  with $k=k_1+k_2$)
\beq
k_\lambda \Delta^{\lambda\mu\nu} (k_1,k_2)= a_n \epsilon^{\mu\nu\alpha\beta}\, k_{1\alpha} \, k_{2\beta}
\eeq
rather trivially, using the longitudinal tensor structure
\beq
\Delta^{\lambda\mu\nu}\equiv w_L= a_n \, \frac{k^{\lambda}}{k^2} \, \epsilon^{\mu\nu\alpha\beta}\, k_{1\alpha} \, k_{2\beta},
\label{IRpole}
\eeq
where $a_n = -i / 2 \pi^2$ denotes the anomaly.
The presence of this tensor structure with a $1/k^2$ behaviour is the signature of the anomaly. 
This result holds for an $AVV$ graph, but can be trivially generalized to more general anomaly graphs, such as 
$AAA$ graphs, by adding poles in the invariants of the remaining lines, i.e. $1/k_1^2$ and $1/k_2^2$, 
by imposing an equal distribution of the anomaly on the three axial-vector legs of the graph.

Obviously, in the chiral limit, the triangle amplitude and the pole amplitude coincide only if the two photons are 
on-shell. In fact, as shown  by Dolgov and Zakharov \cite{Dolgov:1971ri}, the pole
dominance requires a special kinematics. For this reason, the pole has a nonvanishing residue only for massless photons. This, in fact, sets a limit on the validity of the matching, since the perturbative correlator and the pole amplitude are not supposed to coincide for any virtuality of the photons. 
\subsection{UV completions and decoupled poles in the IR}
Being the anomaly closely related to the presence of a pole in the correlation function, the subtraction of the anomaly pole from the perturbative amplitude is sufficient to restore the Ward identities of the theory. For this to occur one has to show that the correlator has always an anomaly pole, which is not obvious. The main goal of this study is to show that the correlator responsible for the chiral gauge anomaly is always (i.e. under any kinematical conditions) characterized by the presence of a pole, and to provide an interpretation of this.  

 We recall that anomaly poles have been identified via an analysis in the IR which shows that the anomalous correlator has indeed a pole characterized by a nonvanishing residue. In fact, the IR coupling of the pole present in the correlator is, for a standard IR pole, rather obvious since the limit
\beq
\lim_{k^2\rightarrow 0} \, k^2 \, \Delta^{\lambda\mu\nu}=k^\lambda \,  a_n \, \epsilon^{\mu\nu\alpha\beta} \, k_{1\alpha} \, k_{2\beta}
\eeq
allows to attribute to the anomaly amplitude a non-vanishing residue. Our main conclusion is that anomaly poles should not be searched for only by the usual dispersive analysis, which is effective only for standard IR poles,  but require a complete off-shell evaluation of the anomalous effective action. We show that these additional poles are decoupled in the IR, but they nevertheless control the UV behaviour of the theory. This last point is proved by looking at a special class of amplitudes which are pole dominated in the UV and which allow to detect the non unitary behaviour of an anomalous theory rather closely.

For this to happen one needs a separation of the anomaly amplitude into longitudinal and transverse components. Our results are based on direct computations, using the two parameterizations of the anomaly amplitude mentioned above. We work under the most general kinematic conditions, generalizing the L/T parameterization given in \cite{Knecht:2003xy} away from the
chiral limit  and showing its exact equivalence to that of Rosenberg.

We start our discussion by addressing the issue of the extraction of an anomaly pole from the Rosenberg form of the anomaly diagram. We review the identification of the independent structures of the AVV diagram in this formulation and then move to the L/T decomposition, illustrating the connection between the two.
\subsection{Connecting two parameterizations}
In his classic paper Rosenberg provided an expression for  the three-point correlator in terms of a sum of six invariant amplitudes multiplied by different tensorial structures, denoted by $A_1, \dots A_6$. These are given as parametric integrals and are easily computable only in few cases, for example when the external momenta are on-shell (massless) or with symmetric off-shell configurations of the two vector lines ($k_1^2=k_2^2$). We will re-analyze the derivation of the amplitude, emphasizing the features of the vertex in the most general case, by focusing our attention on the special kinematical limits in which the pole appears.
\begin{figure}[t]
\begin{center}
\includegraphics[width=15cm]{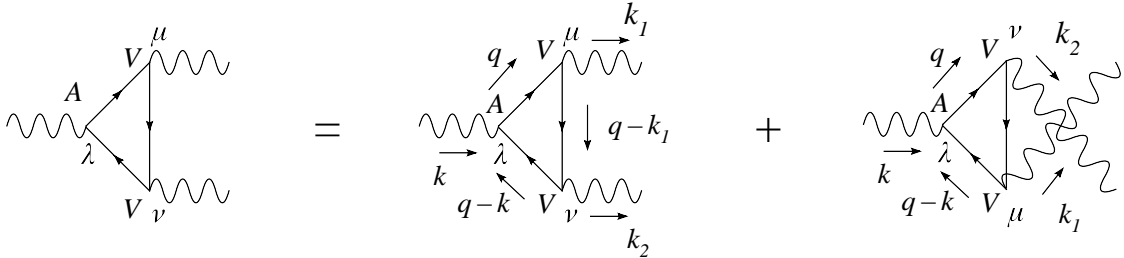}
\caption{\small Triangle diagram with an axial-vector current ($\la$) and two vector currents ($\mu$, $\nu$). The momentum parameterization for the direct and the exchange contribution is written here in an explicit form for future reference.}
\label{AVV}
\end{center}
\end{figure}
The $AVV$ amplitude with off-shell external lines shown in Fig.\ref{AVV} is therefore written according to
\cite{Rosenberg:1962pp}  in the form
\bea
\Delta_0^{\la \mu \nu }=  \frac{i ^3}{(2 \pi )^4} \int  d^4 q
\frac{ Tr \left[
\gamma^{\lambda }\gamma^{5} ( \ds{q} - \ds{k})  \gamma^{\nu} (\ds{q} - {\ds{k}_1}) \gamma^{\mu} \ds{q} \right] }
{ q^2 \, (q - k)^2  \, ( q - k_1 )^2   } \mbox{+ exch.}
\label{delta}
\eea
with
\bea
\Delta_0^{\la\mu\nu} &=& A_1 (k_1, k_2) \veps [k_1,\mu,\nu,\la] + A_2 (k_1, k_2)\veps [k_2,\mu,\nu,\la] +
A_3 (k_1, k_2) \veps [k_1,k_2,\mu,\la]{k_1}^{\nu} \nonumber \\
&+&  A_4 (k_1, k_2) \veps [k_1,k_2,\mu,\la]k_2^{\nu}
+ A_5 (k_1, k_2)\veps [k_1,k_2,\nu,\la]k_1^\mu
+ A_6 (k_1, k_2) \veps [k_1,k_2,\nu,\la]k_2^\mu.\nonumber \\
\label{Ros}
\eea
 The four invariant amplitudes $A_i$ for $i\geq3$ are finite and given by explicit parametric integrals \cite{Rosenberg:1962pp}
\bea
A_3(k_1, k_2) &=& - A_6 (k_2, k_1) =  - 16 \pi^2 I_{11}(k_1, k_2), \\
A_4(k_1,k_2) &=& - A_5 (k_2, k_1) = 16 \pi^2 \left[ I_{20}(k_1,k_2) - I_{10}(k_1,k_2) \right],
\eea
where the general massive $I_{st}$ integral is defined by
\bea
I_{st}(k_1,k_2) = \int_0^1 dw \int_0^{1-w} dz w^s z^t \left[ z(1-z) k_1^2 + w(1-w) k_2^2 + 2 w z (k_1 k_2) - m^2 \right]^{-1},
\eea
whose explicit form will be worked out below.
Both $A_1$ and $A_2$ are instead represented by formally divergent integrals, which can be rendered finite only by imposing the Ward identities on the two vector lines, giving
\bea
A_1 (k_1,k_2) &=& k_1 \cdot k_2 \, A_3 (k_1,k_2) + k_2^2 \, A_4 (k_1,k_2),
\label{WI1} \\
A_2 (k_1,k_2) &=& k_1^2 \, A_5 (k_1,k_2) + k_1 \cdot k_2 \, A_6 (k_1,k_2),
\label{WI2}
\eea
which allow to re-express the formally divergent amplitudes in terms of the convergent ones.
The  Bose symmetry on the two vector vertices with indices $\mu$ and $\nu$ is fulfilled thanks to the relations
\bea
A_5(k_1,k_2) &=& - A_4(k_2, k_1)\\
A_6(k_1, k_2) &=& - A_3 (k_2, k_1).
\eea

\subsection{Explicit expressions in the massless case}
\label{regions}
To extract the explicit form of the parametric integrals given by Rosenberg, we proceed with a direct computation of the invariant amplitudes of the parameterization using dimensional reduction. We perform the traces in 4 dimensions and the loop tensor integrals in $D$ dimensions, using the common techniques of tensor reduction. We use dimensional regularization with minimal subtraction and find, as expected, the cancellation of the dependence of the result on the renormalization scale. Therefore, the parametric integral $I_{11}$ and the combinations $I_{20} - I_{10}$ are trivially identified at the end of the computation. The result is expressed in terms of elementary functions, except for  the function $\Phi(x,y)$
\cite{Usyukina:1993ch}, which is related to one of the  two master integrals of the decomposition, the scalar massless triangle. We obtain for generic virtualities of the external lines
\bea
A_1(s, s_1, s_2) &=&-\frac{i}{4 \pi ^2}
+\frac{i}{8 \pi ^2 \sigma}
\left\{
\Phi(s_1,s_2) \frac{s_1 s_2
   \left(s_2-s_1\right)}{s}
   + s_1\left(s_2-s_{12}\right) \log
   \left[\frac{s_1}{s}\right] \right. \nonumber \\
 && \left.\hspace{3cm}  -s_2\left(s_1-s_{12}\right) \log
   \left[\frac{s_2}{s}\right]
\right\}  ,
 \label{A1massless} \\
A_3 (s, s_1, s_2) &=&
\frac{i}{8 \pi ^2 s \sigma ^2}
\left\{- s_1 s_2 \left[ 4 s_{12}^ 2 + 3  \left(s_1+s_2\right) s_{12}+2 s_1 s_2\right]  \Phi(s_1,s_2)  \right. \nn \\
&& \hspace{3cm}  -2 s   s_{12} \sigma - s s_1 \left[ 2 s_1 s_2+s_{12} \left(3
   s_2+s_{12}\right)\right] \log  \left[ \frac{s_1}{s}\right]  \nn \\
   && \left. \hspace{3cm}  - s s_2
   \left[ s_{12}^2+s_1 \left(2 s_2 + 3  s_{12}\right)\right] \log  \left[\frac{s_2}{s}\right] \right\},
 \label{A3massless} \\
A_4 (s, s_1, s_2) &=&
\frac{i}{8 \pi ^2 s \sigma ^2}
\left\{s_1 \left[ 4 s_{12}^3+2    \left(s_1+2 s_2\right) s_{12}^2+2 s_1 s_2  s_{12}+s_1 \left(s_1-s_2\right) s_2\right] \Phi(s_1,s_2) \right. \nn \\
    && \hspace{3cm} +2 s s_1 \sigma  +s \left(s_1+s_{12}\right) \left(2 s_{12}^2+s_1 s_2\right) \log \left[\frac{s_2}{s}\right] \nn \\
    && \left. \hspace{3cm}  + s s_1   \left[4 s_{12}^2-s_1 \left(s_2-3  s_{12}\right)\right] \log
   \left[\frac{s_1}{s}\right]\right\},
\label{A4massless}
\eea
where $s = k^2$, $s_1= k_1^2  $, $s_2 = k_2^2 $, $s_{12} = k_1 \cdot k_2$
with $\sigma = s_{12}^2 - s_1 s_2$   and the function $\Phi (x, y)$  is defined as
\cite{Usyukina:1993ch}
\bea
\Phi( x, y) &=& \frac{1}{\la} \biggl\{ 2 [Li_2(-\rho  x) + Li_2(- \rho y)]  +
\ln \frac{y}{ x}\ln \frac{1+ \rho y }{1 + \rho x}+ \ln (\rho x) \ln (\rho  y) + \frac{\pi^2}{3} \biggr\},
\label{Phi}
\eea
with
\bea
 \la(x,y) = \sqrt {\Delta},
 \qquad  \qquad \Delta=(1-  x- y)^2 - 4  x  y,
\label{lambda} \\
\rho( x,y) = 2 (1-  x-  y+\la)^{-1},
  \qquad  \qquad x=\frac{s_1}{s} \, ,\qquad \qquad y= \frac {s_2}{s}\, .
\eea
 $\Phi(x,y)$ can be traced back to the one-loop three-point massless scalar integral $C_0(s,s_1,s_2)$, as mentioned above, involved in the reduction of the tensor integrals with three denominators in Eq.~(\ref{delta}) as
\bea
C_0(s,s_1,s_2) = \frac{ i \pi^2}{s} \Phi(x,y).
\eea
Each term in the function $\Phi(x,y)$ and also the arguments of the logarithmic functions appearing in the form factors $A_i$ ($i=1,\dots, 6$) are real if one of these two sets of different conditions is simultaneously satisfied. In the spacelike region we may have
\begin{itemize}
\item $s, s_1, s_2 < 0 $ \qquad and \qquad $s < - (\sqrt{- s_1} + \sqrt{-s_2})^2$
\end{itemize}
or in the physical region with positive kinematical invariants
\begin{itemize}
\item $s, s_1, s_2 > 0 $ \qquad and \qquad $s >  (\sqrt{ s_1} + \sqrt{s_2})^2$.
\end{itemize}
All the other regions would require some specific analytic continuations by giving to all the invariants a small imaginary part $\eta$ ($\eta>0$) according to the $i \eta$ prescription with $s_i \to s_i + i \eta$.

When discussing the presence of spurious poles for $s \to 0$ we need to work with amplitudes which are well-defined  around $s=0$; for this reason the analytic regularizations have been  always performed  before taking the $s \to 0$ limit. There is another important observation that is in order at this point. One may worry if the absence of the pole in $s$ can be attributed to the redundancy of the Rosenberg representation, but, as we are going to show next, this is not the case.
\subsection{Four amplitude decomposition in Rosenberg}
In order to derive a set of a minimal number of independent invariant amplitudes we proceed from scratch. The identification of the invariant tensor structures characterizing the amplitude can be done exhaustively, by starting with the construction of all the possible tensors of rank three built out of the $\veps$-tensor and the external momenta.
We follow here an approach similar to \cite{Giannotti:2008cv} with some minor changes.  \\
The eight tensorial structures listed in Tab.\ref{table1} are the ones needed in the expansion of a generic triangle correlator with three indices $\{ \la, \mu, \nu\}$ and external momenta $\{k_1, k_2\}.$ Out of these 8 structures, only the six in the first three columns  appear in the Rosenberg formulation and can be reduced to 4  with little effort by requiring conservation of the vector currents. If we impose the vector Ward identity on the two vector lines of the diagram and fix the divergent coefficients $A_1$ and $A_2$ in terms of the remaining amplitudes, then the form factors $A_i$ reduce to the four ones $A_3, \dots, A_6$ and the tensor structures in front of them get automatically organized in terms of four linear combinations indicated with $\eta_i$. These four tensor amplitudes $\eta_i$ are selected from a set of six quantities defined in Tab.\ref{table2}, which shows all the possible tensors entering into the expansion of a generic three-currents correlator \emph{after} imposing the conservation of the vector current.
\begin{table}[t]
\begin{center}
\begin{tabular}
{|c  c  c c |} \hline
$\veps[k_1,\la, \mu, \nu]   $ \qquad \qquad  &
$\veps[k_1,k_2,\mu, \la] \, k_1^{\nu} $ \qquad \qquad  &
$\veps [k_1,k_2,\nu, \la] \, k_1^{\mu}$ \qquad \qquad  &
$\veps [k_1,k_2,\mu, \nu] \, k_1^{\la}$ \\
$\veps[k_2, \la, \mu, \nu]  $\qquad \qquad &
$\veps[k_1,k_2,\mu, \la] \, k_2^{\nu} $ \qquad \qquad &
$\veps [k_1,k_2,\nu, \la] \, k_2^{\mu}$ \qquad \qquad &
$\veps [k_1,k_2,\mu, \nu] \, k_2^{\la}$ \\
\hline
\end{tabular}
\caption{ The eight pseudotensors in which a general amplitude $\D ^{\l \mu \nu}(k_1, k_2) $ can be expanded.
\label{table1} }
\end{center}
\end{table}

\begin{table}[t]
\begin{center}
\begin{tabular}
{|c |c |}   \hline
$\eta_1$ & $\veps [k_1,k_2,\mu, \nu] \, k_1^{\la} $  \\ \hline
$\eta_2$ & $\veps [k_1,k_2,\mu, \nu] \, k_2^{\la}$ \\ \hline\hline
$\eta_3$         &           $k_1 \cdot k_2 \varepsilon[k_1,\la, \mu, \nu] + k_1^{\nu} \varepsilon[k_1,k_2,\mu, \la]$ \\ \hline
$\eta_4$ & $k_2 \cdot k_2 \varepsilon[k_1,\la, \mu, \nu] + k_2^{\nu} \varepsilon[k_1,k_2,\mu, \la] $ \\ \hline
$\eta_5$ & $ k_1 \cdot k_1 \varepsilon[k_2,\la, \mu, \nu] + k_1^{\mu} \varepsilon[k_1,k_2,\nu, \la] $  \\ \hline
$\eta_6$ & $k_1 \cdot k_2 \varepsilon[k_2,\la, \mu, \nu] + k_2^{\mu} \varepsilon[k_1,k_2,\nu, \la]  $  \\ \hline
\end{tabular}
\caption{ The six pseudotensors needed in the expansion of an amplitude $\D ^{\l \mu \nu}(k_1, k_2) $ satisfying the vector current conservation. \label{table2} }
\end{center}
\end{table}

Coming back to our specific case, we obtain for the generic anomalous $AVV$ vertex  satisfying the vector Ward identities the parameterization
\bea
\Delta^{\la \mu \nu}_{WI} &=& A_3 (k_1 \cdot k_2 \varepsilon[k_1,\la, \mu, \nu] + k_1^{\nu} \varepsilon[k_1,k_2,\mu, \la]) +
A_4 (k_2 \cdot k_2 \varepsilon[k_1,\la, \mu, \nu] + k_2^{\nu} \varepsilon[k_1,k_2,\mu, \la] ) \nn \\
&& + A_5 (k_1 \cdot k_1 \varepsilon[k_2,\la, \mu, \nu] + k_1^{\mu} \varepsilon[k_1,k_2,\nu, \la]) +
A_6 (k_1 \cdot k_2 \varepsilon[k_2,\la, \mu, \nu] + k_2^{\mu} \varepsilon[k_1,k_2,\nu, \la] ) \nn\\
&& = A_3 \, \eta_3^{\la\mu\nu}(k_1,k_2) + A_4 \, \eta_4^{\la\mu\nu}(k_1,k_2)
+ A_5 \, \eta_5^{\la\mu\nu}(k_1,k_2) + A_6 \, \eta_6^{\la\mu\nu}(k_1,k_2).
\label{reduced}
\eea
This is obtained after plugging Eqs.~(\ref{WI1},\ref{WI2})  into Eq.(\ref{Ros}), where $ \eta_i^{\la\mu\nu}(k_1,k_2)$  can be read from Tab.\ref{table2}.
The remaining two homogeneous pseudotensors of degree $3$ in $k_1,k_2$, denoted by  $ \eta_1^{\la\mu\nu}$ and $ \eta_2^{\la\mu\nu}$
\bea
\eta_1^{\la\mu\nu}(k_1,k_2) =  k_1 ^{\la} \, \varepsilon [k_1,k_2,\mu, \nu],  \qquad \qquad
\eta_2^{\la\mu\nu}(k_1,k_2) =  k_2 ^{\la} \, \varepsilon [k_1,k_2,\mu, \nu],
\eea
are not present in the Rosenberg parameterization, although they appear in the L/T decomposition, as we show below. The reduction of these two tensors
 to the four ones already used as a basis can be achieved by the use of two Schouten relations
\bea
k_1^ {\la}\, \veps[k_1,k_2,\mu, \nu]  &= &
k_1^ {\mu} \, \veps[k_1,k_2,\la, \nu] - k_1^\nu \veps[k_1,k_2,\la, \mu] - \,
k_1^2  \veps[k_2,\la,\mu, \nu] + k_1 \cdot k_2  \veps[k_1,\la,\mu, \nu], \nn \\
\label{schouten1} \\
k_2^ \la \, \veps[k_1,k_2,\mu, \nu]  &= &
k_2^\mu \,  \veps[k_1,k_2,\la, \nu] - k_2^\nu \, \veps [k_1,k_2,\la, \mu]  - \,
k_1 \cdot k_2 \veps [k_2,\la,\mu, \nu] + k_2^2 \veps [k_1,\la,\mu, \nu], \nn \\
\label{schouten2}
\eea
or equivalently,
\bea
\eta_1 ^{\la\mu\nu}(k_1,k_2) &=& \eta_3 ^{\la\mu\nu}(k_1,k_2) - \eta_5 ^{\la\mu\nu}(k_1,k_2),    \\
\eta_2 ^{\la\mu\nu}(k_1,k_2) &=& \eta_4 ^{\la\mu\nu}(k_1,k_2) - \eta_6 ^{\la\mu\nu}(k_1,k_2).
\eea
The set of the 4 amplitudes that we have chosen in the parameterization shown in Eq.~(\ref{reduced}) are linearly independent and functionally independent respect to the Schouten transformations. The claim that one can make is that any tensor structure which is not of the form given in the 4-basis above can be re-expressed as a combination of these 4 structures using appropriate Schouten relations. The decomposition of the AVV diagram with respect to this basis is therefore unique. At this point it is trivial to realize that, starting from the explicit expressions of the invariant amplitudes $A_i$ that we have given above, the absence of a residue at $s=0$ continues to hold (for general off-shell kinematics). The important point to observe is that there is no kinematical singularity in this limit in each of the 4 independent tensor structures. The conclusion is that, in general, an AVV diagram has no massless poles. The use of a set of non-redundant amplitudes clears the ground of any doubt concerning this result.  In fact, the poles appear only under special kinematical configurations, as we are going to discuss next.

\section{The massive off-shell case for the Rosenberg parameterization }
Before performing the relevant kinematical limits on the amplitude, we move one step forward and generalize the results presented in the previous section to the massive case, by writing the expression of the invariant amplitudes given by Rosenberg (and the corresponding parametric integrals) in an explicit form.

The computation is performed as in the massless case, using dimensional reduction. The modifications are minimal and mostly due to the new scalar integrals $B_0$ and $C_0$, corresponding to the massive (scalar) self-energy and triangle diagram respectively.
The three-point amplitude with equal massive internal lines is given by
\bea
\D^{\l \mu \nu }=  \frac{i ^3}{(2 \pi )^4} \int  d^4 q
\frac{ Tr \left[
\gamma^{\lambda }\gamma^{5} ( \ds{q} - \ds{k} + m)  \gamma^{\nu} (\ds{q} - {\ds{k}_1}+m ) \gamma^{\mu} (\ds{q}+m) \right] }
{ (q^2- m^2) \, ((q - k)^2 -m^2) \, ( (q - k_1 )^2-m^2)   } \mbox{+ exch.},
\eea
\label{deltam}
with $k = k_1+k_2 $, and can be again cast into the form
\bea
\Delta^{\la\mu\nu} &=&
A_1 (k_1, k_2, m^2) \, \veps [k_1,\mu,\nu,\la] + A_2 (k_1, k_2, m^2)\, \veps [k_2,\mu,\nu,\la]  \nn \\
&+& A_3 (k_1, k_2, m^2) \, \veps [k_1,k_2,\mu,\la]\, {k_1}^{\nu}
+  A_4 (k_1, k_2, m^2) \, \veps [k_1,k_2,\mu,\la]\, k_2^{\nu} \nn \\
&+& A_5 (k_1, k_2, m^2) \, \veps [k_1,k_2,\nu,\la] \, k_1^\mu
+ A_6 (k_1, k_2, m^2) \, \veps [k_1,k_2,\nu,\la] \, k_2^\mu,
\eea
where the tensorial structures are the same as before and the massive form factors $A_i (k_1, k_2, m^2)$ show an explicit dependence on the internal mass. They have been computed by using the tensor reduction technique to express the tensorial one-loop integrals in terms of the scalar ones. We obtain
\bea
A_1(k_1, k_2, m^2) &=&
 - \frac{i}{4 \pi^2 }  + \frac{1}{8 \pi^4 \sigma} \, \left\{ s_1  \left(s_2-s_{12}\right)
   D_1\left(s_1,s,m^2\right) -s_2  \left(s_1-s_{12}\right) D_2\left(s_2,s,m^2\right) \right. \nn \\
    &+& \left. \left[ s_1 s_2 \left(s_2-s_1\right) - 4 \sigma m^2  \right]
   C_0\left(s_1,s_2,s,m^2\right)  \right\},
   \label{A1mass}\\
A_3(k_1, k_2, m^2) &=&
- \frac{i}{4 \pi^2 \sigma} s_{12} + \frac{1}{8 \pi^4 \sigma^2}\,
\left \{ -s_1 \left[2 s_1 s_2+s_{12} \left(3 s_2+s_{12}\right)\right] \, D_1\left(s_1,s,m^2\right) \right. \nn \\
&-& s_2 \left[ 2 s_1 s_2+s_{12} \left(3 s_1+s_{12}\right)\right] \, D_2\left(s_2,s,m^2\right) \nn \\
 &-& \left.    \left[    4 s_{12} \sigma m^2   +  s_1 s_2 \left(4 s_{12}^2+3  \left(s_1+s_2\right) s_{12}+2 s_1 s_2\right)  \right] \,
   C_0\left(s_1,s_2,s,m^2\right)\right \} ,
\label{A3mass} \\
A_5 (k_1, k_2, m^2) &=&
- \frac{i}{4 \pi^2 \sigma} s_{2} + \frac{1}{8 \pi^4 \sigma^2}
\left\{ -\left(s_2+s_{12}\right)
   \left(2 s_{12}^2+s_1 s_2\right)
   D_1\left(s_1,s,m^2\right)\right. \nn \\
   &-& s_2 \left[ s_{12} \left(3s_2+4 s_{12}\right)-s_1 s_2\right] \, D_2\left(s_2,s,m^2\right) \nn \\
   &-&  \left. \left[ 4 s_2 \sigma m^2   +  s_2 \left(-s_2  s_1^2+\left(s_2^2+2 s_{12} s_2+4 s_{12}^2\right)
   s_1   \right. \right.\right. \nn \\
    && \left. \left. \left. +2 s_{12}^2 \left(s_2+2
   s_{12}\right)\right)\right] \,
   C_0\left(s_1,s_2,s,m^2\right)\right\},
   \label{A5mass}
\eea
with $s = k^2$, $s_1= k_1^2  $, $s_2 = k_2^2 $,  $\sigma = s_{12}^2 - s_1 s_2$. It is possible to check that the Bose symmetry relative to the two vector vertices
\bea
A_2 (k_1, k_2, m^2) &=& - A_1 (k_2, k_1, m^2), \\
A_6 (k_1, k_2, m^2) &=& - A_3 (k_2, k_1, m^2), \\
A_4 (k_1, k_2, m^2) &=& - A_5 (k_2, k_1, m^2)
\eea
is respected.
As mentioned above, the difference between the massless and the massive decomposition of the triangle amplitude lies in the particular set of scalar integrals involved in the tensor reduction. Here we define $D_1$ and $D_2$ as a combination of two-point scalar massive integrals $(B_0)$ of different internal momenta
\bea
D_i (s, s_i,  m^2)&=& B_0 (k^2, m^2) - B_0 (k_i^2, m^2) = i \pi^2 \left[ a_i \log\frac{a_i +1}{a_i - 1} - a_3 \log \frac{a_3 +1}{a_3 - 1}  \right] \qquad i=1,2
\label{D_i}
\nn \\
\eea
in which the dependence on the regularization scheme disappears in the difference of the two scalar self-energies involved in (\ref{D_i}). The expression of $C_0$ can be given explicitly in various forms \cite{Kniehl:1989qu}, for instance as
\bea
C_0 (s, s_1, s_2, m^2)&=& - i \pi^2 \frac{1}{2 \sqrt \sigma} \sum_{i=1}^3 \left[Li_2 \frac{b_i -1}{a_i + b_i}   - Li_2 \frac{- b_i -1}{a_i - b_i} +
Li_2 \frac{-b_i +1}{a_i - b_i}  - Li_2 \frac{b_i +1}{a_i + b_i}
   \right]\nn \\
\label{C0polylog}
\eea
with
\bea
a_i = \sqrt {1- \frac{4 m^2}{s_i }}, \qquad \qquad
b_i = \frac{- s_i + s_j + s_k }{2 \sigma},
\eea
where $s_3=s$ and in the last equation $i=1,2,3$ and $j, k\neq i$.  Other expressions, suitable for numerical implementations, are given in \cite{vanOldenborgh:1989wn}.
The region in which all these functions have real arguments and do not need any analytic continuations are those discussed in section \ref{regions}, for the massless case. In general, the prescription for $i \eta$  in the presence of  a mass in the internal loop - in the fermion propagator -  is taken as $m \to m - i \eta$.  We have checked numerically the agreement between the expressions presented above and those given in parametric form.

\section{The vertex in the longitudinal/transverse (L/T) formulation and comparisons}
 The second parameterization of the three-point correlator function that we are going to discuss is the one presented in \cite{Knecht:2003xy}.
 One of the features of this parameterization is the presence of a longitudinal contribution for generic virtualities of the external momenta and not just in the specific configuration under which it appears in Rosenberg's formulation. Of course, the true presence of the pole in the IR has to be checked by taking the corresponding limit, since the Schouten relations allow the extraction of a pole in the IR region at the cost of extra singularities in the parameterization. For this reason we start by recalling the structure of the L/T parameterization, which separates the longitudinal from the transverse components of the anomaly vertex, which is given by

\beq
 \mathcal \, W ^{\lambda\mu\nu}= \frac{1}{8\pi^2} \left [  \mathcal \, W^{L\, \lambda\mu\nu} -  \mathcal \, W^{T\, \lambda\mu\nu} \right],
\label{long}
\eeq
where the longitudinal component
\beq
 \mathcal \, W^{L\, \lambda\mu\nu}= w_L  \, k^\lambda \veps[\mu,\nu,k_1,k_2]
\eeq
(with $w_L=- 4 i /s $) describes the anomaly pole, while the transverse contributions take the form
\beqa
\label{calw}
{  \mathcal \, W^{T}}_{\lambda\mu\nu}(k_1,k_2) &=&
w_T^{(+)}\left(k^2, k_1^2, k_2^2 \right)\,t^{(+)}_{\lambda\mu\nu}(k_1,k_2)
 +\,w_T^{(-)}\left(k^2, k_1^2,k_2^2\right)\,t^{(-)}_{\lambda\mu\nu}(k_1,k_2) \nonumber \\
 && +\,\, {\widetilde{w}}_T^{(-)}\left(k^2, k_1^2, k_2^2 \right)\,{\widetilde{t}}^{(-)}_{\lambda\mu\nu}(k_1,k_2),
 \eeqa
with the transverse tensors given by
\beqa
t^{(+)}_{\lambda\mu\nu}(k_1,k_2) &=&
k_{1\nu}\, \veps[ \mu,\la, k_1,k_2]  \,-\,
k_{2\mu}\,\veps [\nu,\la, k_1, k_2]  \,-\, (k_{1} \cdot k_2)\,\veps[\mu,\nu,\la,(k_1 - k_2)]
\nonumber\\
&& \quad\quad+ \, \frac{k_1^2 + k_2^2 - k^2}{k^2}\, \, k_\la \, \,
\veps[\mu, \nu, k_1, k_2]
\nonumber \ , \\
t^{(-)}_{\lambda\mu\nu}(k_1,k_2) &=& \left[ (k_1 - k_2)_\la \,-\, \frac{k_1^2 - k_2^2}{k^2}\,\, k_\la \right] \,\veps[\mu, \nu, k_1, k_2]
\nonumber\\
{\widetilde{t}}^{(-)}_{\lambda\mu\nu}(k_1,k_2) &=& k_{1\nu}\,\veps[ \mu,\la, k_1,k_2] \,+\,
k_{2\mu}\,\veps [\nu,\la, k_1, k_2] \,
-\, (k_{1}\cdot k_2)\,\veps[ \mu, \nu, \la, k].
\label{tensors}
\eeqa
The form factors $w_T (s, s_1, s_2)$ are all defined in the following Eqs.(\ref{wTpex}-\ref{wTtex}).

Notice that in this representation the presence of massless poles is explicit for any kinematical configuration
and not just in the massless collinear limit, where the diagram takes the Dolgov-Zakharov form. A second observation concerns the presence of other pole-like singularities in the transverse invariant amplitude and tensor structures. It is then obvious that one has to wonder whether the pole present in $w_L$ is balanced, away from the collinear region, by other contributions which are also singular.  Indeed, as we are going to show, this is the case. In fact, due to the Schouten relations, we are always allowed to introduce new polar amplitudes and balance them with additional contributions on the remaining tensor structures. In fact we are going to show that the presence of such pole away from the collinear region becomes significant in the UV - at least in the perturbative approach - but not in the IR, since it decouples if one computes the residue correctly in this representation.

\subsection{Generalizing the L/T parameterization to massive fermions and the anomaly pole}
We can generalize the L/T formulation presented above to the case of a triangle amplitude with  a massive fermion of mass $m$, by simply exploiting the connection between this and the Rosenberg representation. We use the Schouten relations to show the equivalence between the tensor structures of both representations. This requires some care since the decomposition into $L$ and $T$ amplitudes requires a nonzero $k$, otherwise it is invalid.

At nonzero momentum, by equating the coefficients of the four invariant tensors, we obtain a linear system of four equations whose solutions return the complete matching between the two parameterizations in the form
\bea
A_3 (k_1, k_2) &=& \frac{1}{8 \pi^2} \left[ w_L - \tilde{w}_T^{(-)}
- \frac{k^2}{(k_1+ k_2)^2}       w_T^{(+)}
- 2 \,  \frac{k_1 \cdot k_2 - k_2^2 }{k^2 }w_T^{(-)}  \right],  \\
A_4 (k_1, k_2) &=& \frac{1}{8 \pi^2} \left[  w_L
+ 2 \, \frac{k_1 \cdot k_2}{k^2}       w_T^{(+)}
- 2 \, \frac{k_1 \cdot k_2 + k_2^2}{k^2 }w_T^{(-)}  \right], \\
A_5 (k_1,k_2) &=& - A_4 (k_2, k_1), \qquad \qquad A_6 (k_1,k_2) = - A_3 (k_2, k_1),
\eea
and viceversa
\bea
w_L (k^2, \, k_1^2, \, k_2^2) &=& \frac{8 \pi^2}{k^2} \left[A_1 - A_2 \right],
\eea
(we omit, for simplicity, the momentum dependence)
or, after the imposition of the Ward identities in Eqs.(\ref{WI1},\ref{WI2}),
\bea
w_L (k^2, \, k_1^2, \, k_2^2) &=& \frac{8 \pi^2}{k^2}
\left[ (A_3-A_6) k_1 \cdot k_2 + A_4 \, k_2^2 - A_5 \, k_1^2 \right],
\label{wL}\\
w_T^{(+)} (k^2, \, k_1^2, \, k_2^2)  &=& - 4 \pi^2 \left(A_3 - A_4 + A_5 - A_6 \right),
\label{wTp}\\
w_T^{(-)} (k^2, \, k_1^2, \, k_2^2)  &=&  4 \pi^2 \left(A_4+A_5 \right),
\label{wTm}\\
\tilde{w}_T^{(-)} (k^2, \, k_1^2, \, k_2^2)  &=& - 4\pi^2 \left( A_3 + A_4 + A_5 + A_6 \right),
\label{wTt}
\eea
where $A_i\equiv A_i(k_1,k_2)$. This same mapping holds also in the massive fermion case if $A_i\equiv A_i(k_1,k_2,m)$ and leads us to the same decomposition. In this case the L/T parameterization can be obtained starting from the massive $A_i$ coefficients shown in Eq.(\ref{A1mass}-\ref{A5mass})  and exploiting the mapping in Eqs.~(\ref{wL}-\ref{wTt}) between the two parameterizations. We obtain
\bea
w_L (s_1,s_2,s) &=& - \frac{4 i}{s} \\
w_T^{(+)}(s_1,s_2,s) &=& i\frac{s}{\sigma} + \frac{i}{2 \sigma^2} \left[
(s_{12}+s_2)(3 s_1^2 + s_1(6 s_{12}+s_2)+2s_{12}^2)\log \frac{s_1}{s} \right. \nn \\
&+& (s_{12}+s_1)(3 s_2^2 + s_2(6 s_{12}+s_1)+2s_{12}^2)\log \frac{s_2}{s} \nn \\
&+& \left. s (2 s_{12}(s_1+s_2)+s_1 s_2(s_1+s_2+6s_{12})) \Phi (s_1,s_2)
\right]
\label{wTpex} \\
w_T^{(-)}(s_1,s_2,s)&=&  i\frac{s_1-s_2}{\sigma} +\frac{i}{2 \sigma^2} \left[
-(2(s_2+s_{12}) s_{12}^2 - s_1 s_{12} (3s_1 + 4 s_{12}) \right. \nn \\
&+& \left. s_1 s_2 (s_1+s_2+s_{12})) \log \frac{s_1}{s}
 + (2(s_1+s_{12}) s_{12}^2 - s_2 s_{12} (3s_2 + 4 s_{12}) \right. \nn \\
 &+& s_1 s_2 (s_1 + s_2 + s_{12})) \log \frac{s_2}{s} + \left. s (s_1-s_2)(s_1 s_2 + 2 s_{12}^2)\Phi (s_1,s_2) \right]
 \label{wTmex}\\
\tilde{w}_T^{(-)}(s_1,s_2,s) &=& - w_T^{(-)}(s_1,s_2,s)
\label{wTtex}
\eea
in the massless case,
which is in complete agreement with the explicit expression given by \cite{Jegerlehner:2005fs}, while in the massive case the same mapping gives
\bea
w_L (s, s_1,s_2,m^2) &=& - \frac{4 i}{s} -\frac{8 m^2}{\pi ^2 s} C_0(s, s_1,s_2,m^2) \\
w_T^{(+)}(s, s_1, s_2, m^2) &=& i\frac{s}{\sigma}+\frac{1}{2 \pi^2 \sigma^2} \left[
(s_{12}+s_2)(3 s_1^2 + s_1(6 s_{12}+s_2)+2s_{12}^2)D_1(s, s_1,m^2) \right. \nn \\
&+& (s_{12}+s_1)(3 s_2^2 + s_2(6 s_{12}+s_1)+2s_{12}^2)D_2(s,s_2,m^2) \nn \\
&+& \left. (4 m^2 s \sigma + s (2 s_{12}(s_1+s_2)+s_1 s_2(s_1+s_2+6s_{12})))C_0(s,s_1,s_2,m^2)
\right] \\
w_T^{(-)}(s, s_1,s_2,m^2)&=&  i\frac{s_1-s_2}{\sigma} +\frac{1}{2 \pi^2 \sigma^2} \left[
-(2(s_2+s_{12}) s_{12}^2 - s_1 s_{12} (3s_1 + 4 s_{12}) \right. \nn \\
&+& \left. s_1 s_2 (s_1+s_2+s_{12})) D_1(s,s_1,m^2)
 + (2(s_1+s_{12}) s_{12}^2 - s_2 s_{12} (3s_2 + 4 s_{12}) \right. \nn \\
 &+&  s_1 s_2 (s_1 + s_2 + s_{12})) D_2(s, s_2,m^2) \nn \\
&+& \left. (4 m^2 \sigma (s_1-s_2) + s (s_1-s_2)(s_1 s_2 + 2 s_{12}^2))C_0(s,s_1,s_2,m^2) \right] \\
\tilde{w}_T^{(-)}(s, s_1,s_2,m^2) &=& - w_T^{(-)}(s,s_1,s_2,m^2),
\eea
with  $s_i = k_i^2$ ($i= 1, 2, 3$, $k_3=k$), $s_{12}= k_1\cdot k_2$, $\sigma = s_{12}^2 - s_1 s_2$. The functions $D_i$ and $C_0$,  defined in Eq.(\ref{D_i}) and (\ref{C0polylog}), are respectively a combination of two scalar bubbles 
and the scalar one-loop triangle.
 The Bose symmetry on the vector vertices is fulfilled in both representations by taking into account the way in which the $A_i$ and the $w_L, w_T, \dots$ transform under the exchange of $k_1, k_2$ and $\mu, \nu$. For the  L/T invariant amplitudes we have
\bea
w_T^{(+)} (k^2, \, k_1^2, \, k_2^2)  &=& w_T^{(+)} (k^2, \, k_1^2, \, k_2^2), \\
w_T^{(-)} (k^2, \, k_1^2, \, k_2^2)  &=& - w_T^{(-)} (k^2, \, k_1^2, \, k_2^2), \\
\tilde{w}_T^{(-)} (k^2, \, k_1^2, \, k_2^2)  &=& - \tilde{w}_T^{(-)} (k^2, \, k_1^2, \, k_2^2).
\eea
It is then obvious that there is complete equivalence between the two parameterizations, although there are some puzzling features that need to be investigated more closely. As we have already mentioned, the L/T parameterization appears to have a pole at
$s=(k_1 +k_2)^2=0$, which contributes to the anomaly. In fact, the non-vanishing Ward identity on the axial-vector line is due to the invariant amplitude $w_L$ and to its corresponding tensor structure.
Then, one obvious question to ask is if this pole is compatible with the pole structure of the Rosenberg representation. The answer is affirmative as far as the computation of the residue is performed on the entire amplitude and not just on the invariant amplitudes alone. In fact, the L/T decomposition introduces kinematical singularities both in the longitudinal and in the transverse components as a price for the appearance of a longitudinal pole. This can be shown explicitly. In fact, a direct evaluation of the limit
(for off shell photons) gives
\bea
&& \lim_{s \rightarrow 0 } \,   s \, w_L(k_1^2, k_2^2, k^2) (k_1+ k_2)_{\la} \veps[\mu, \nu, k_1, k_2] =
- 4 i  (k_1+ k_2)_{\la} \veps[\mu, \nu, k_1, k_2],\\
&& \lim_{s \rightarrow 0 } \,   s \, w_T^{(+)} (k_1^2, k_2^2, k^2) \,  t _{\mu \nu \la} ^{(+)}(k_1, k_2)  =
-\frac{2 i (s_1+s_2)  \log  [\frac{s_1}{s_2}]   }  {{s_1}-{s_2}}  (k_1+ k_2)_{\la} \veps[\mu, \nu, k_1, k_2], \\
&& \lim_{s \rightarrow 0 } \,   s \,  w_T^{(-)} (k_1^2, k_2^2, k^2) \, t _{\mu \nu \la} ^{(-)}(k_1, k_2) =
\left [- 4 i + \frac{2 i (s_1+s_2)  \log  (\frac{s_1}{s_2})   }  {{s_1}-{s_2}} \right]  (k_1+ k_2)_{\la} \veps[\mu, \nu, k_1, k_2], \nn \\ \\
&& \lim_{s \rightarrow 0 } \,   s \, \tilde w_T^{(-)} (k_1^2, k_2^2, k^2) \, \tilde t _{\mu \nu \la} ^{(-)}(k_1, k_2) = 0
\eea
for the several singular terms present at $s=0$. These results have been obtained after performing the analytic continuation around $s=0$ of the explicit expressions for $w_L$ and $w_T $ given above. Combining these partial contributions we obtain the total result for the residue of the entire amplitude
\beq
\lim _{s \rightarrow 0 } \,  s \mathcal \, W _{\mu \nu \la}=0,
\label{limit}
\eeq
which proves its vanishing at $s=0$ for off-shell photon lines. This result, in agreement with what we had anticipated, shows that in the IR also the L/T parameterization has no pole. This is expected, being the L/T and the Rosenberg parameterizations equivalent descriptions
of the same diagram (modulo some Schouten relations), hence it is obvious that the decoupling of the anomaly
pole for off-shell external momenta has to take place in both parameterizations. Performing cautiously the limits, we can similarly proof that the pole
reappears in correspondence of specific configurations of the external lines (on-shell photons), as we are going to show next. An equivalent analysis, of course, can be performed by
analyzing the various cuts of the amplitudes in the L/T parameterization using a dispersive approach and looking for discontinuities proportional to $\delta(k^2)$ in the spectral density of the diagram.

\section{Special kinematical limits in the massless case}
We summarize in this section all the results concerning some specific kinematical conditions in the infrared and chiral limits of the anomaly amplitude, taken directly on the amplitude given in the previous sections.

The first analysis carried out involves the massless $A_i$ written in Eq.(\ref{A1massless}, \ref{A4massless}) for which we take  three limits.  We use the notation $A_i(s, s_1,s_2)$ to denote each invariant amplitude in the Rosenberg form for massless internal fermions. We distinguish the following cases
\begin {itemize}
\item  [a)]  $s_1 =0$ \qquad $s_2 \neq 0$ \qquad $s \neq 0$ \qquad $m = 0$
\item[b)] $s_1 =0$ \qquad $s_2 = 0$ \qquad $s \neq 0$ \qquad $m =  0$
\item[c)] $s_1 = M^2$ \qquad $s_2 = M^2$ \qquad $s \neq 0$ \qquad $m =  0$.
\end{itemize}
While cases a) and b) will be treated here, case c) will be left to the appendix \ref{poles_residui}, together with the same three kinematical configurations for a massive fermion.
In case a) we find
\bea
A_1 (s, 0, s_2) &=& \frac{i}{4 \pi^2}\left[ \frac{s_2}{s-s_2}\log\frac{s_2}{s} -1 \right], \\
A_2 (s, 0, s_2) &=& \frac{i}{4 \pi^2}\left[ \frac{s_2}{s-s_2}\log\frac{s_2}{s} +1 \right], \\
A_3 (s, 0, s_2) &=& -A_6 (0, s_2, s, 0)=-\frac{i}{2 \pi^2 (s-s_2)}\left[ \frac{s_2}{s-s_2}\log\frac{s_2}{s} +1 \right], \\
A_4 (s, 0, s_2) &=& \frac{i}{2 \pi^2 (s-s_2)}\log\frac{s_2}{s}
\eea
and a divergent $A_5 (s, 0, s_2)$ which does not contribute to the physical value of the amplitude. Indeed $\Delta^{\la\mu\nu} $, in a physical amplitude, is contracted with the polarization vector relative to the on-shell photon with momentum $k_1$, giving \mbox{$\eps_\mu(k_1) k_1^{\mu}=0$}, so that the contribution coming from $A_5$ disappears. \\
Notice that this amplitude satisfies the Ward identities in Eqs.~(\ref{WI1},\ref{WI2}) and can be written as
\bea
\Delta^{\la \mu \nu} (s, 0, s_2) = A_3 (s, 0, s_2) \, \eta_3 ^{\la \mu \nu}  (k_1, k_2) +
A_4 (s, 0, s_2) \, \eta_4 ^{\la \mu \nu}  (k_1, k_2)  + A_6 (s, 0, s_2) \, \eta_6 ^{\la \mu \nu}  (k_1, k_2), \nn \\
\eea
with the tensors $\eta_i (k_1,k_2)$ written in Tab.\ref{table2}.
Notice that the poles are located at the various thresholds of the amplitude, describing the production of a photon of invariant mass
$s_2$, having set the first photon on-shell, and that all the residues are vanishing
\bea
&& \lim_{s \rightarrow 0 } \,   s \, A_3 (s, 0, s_2)
= \lim_{s \rightarrow 0 } \,   s \, A_4 (s, 0, s_2)
=\lim_{s \rightarrow 0 } \,   s \, A_6 (s, 0, s_2) =0,
\eea
including the one of the whole amplitude
\bea
\lim_{s \rightarrow 0 } \,   s \, \Delta^{\la \mu \nu} (s, 0, s_2) = 0.
\eea

In the L/T parameterization we find
\bea
w_L(s, 0,s_2) &=& - \frac{4i}{s} ,\\
w_T^{(+)}(s, 0,s_2) &=& \frac{2 i}{s-s_2}\left[ \frac{s+s_2}{s-s_2}\log\frac{s_2}{s} +2 \right],\\
w_T^{(-)}(s, 0,s_2) &=& -\tilde{w}_T^{(-)}(s, 0,s_2) = \frac{2 i}{s-s_2}\log\frac{s_2}{s}
\eea
which also show the presence of the same threshold singularity, but, in addition, also of an anomaly pole
in $w_L$ which is absent in Rosenberg's parameterization. As we have commented above, the pole is spurious, since
the tensor structures are also singular in the same ($s\to 0$) limit, and there is a trivial cancellation of this contribution. Indeed we find
\bea
&& \lim_{s \rightarrow 0 } \,   s \, w_L(s, 0, s_2) \, k_{\la} \veps[\mu, \nu, k_1, k_2] =
- 4 i \,  k_{\la} \, \veps[\mu, \nu, k_1, k_2], \\
&& \lim_{s \rightarrow 0 } \,   s \,\left[  w_T^{(+)} (s, 0, s_2) \,  t _{\la \mu \nu} ^{(+)}(k_1, k_2)  +
  w_T^{(-)} (k_1^2, k_2^2, k^2) \, t _{\la \mu \nu} ^{(-)}(k_1, k_2) \right] = - 4 i \,  k_{\la} \, \veps[\mu, \nu, k_1, k_2],  \nn \\ \\
&& \lim_{s \rightarrow 0 } \,   s \, \tilde w_T^{(-)} (s, 0, s_2) \, \tilde t _{\la \mu \nu} ^{(-)}(k_1, k_2) = 0
\eea
which gives
\bea
\lim _{s \rightarrow 0 } \,  s \mathcal \, W _{\la \mu \nu } (s, 0, s_2) =
 \frac{1}{8\pi^2}  \lim _{s \rightarrow 0 }  \,  s \, \left [  \mathcal \, W^{L\, \lambda\mu\nu} -  \mathcal \, W^{T\, \lambda\mu\nu} \right] = 0
\eea
in agreement with Eq. \ref{limit}. \\
Therefore, in this case, with only one leg on-shell, the kinematics does not allow a polar structure for the entire amplitude; in the Rosenberg parameterization this result can be derived in a straightforward way since each amplitude has a vanishing residue and the tensor structures are regular in the IR (i.e. $s\to 0$) limit. On the contrary, in this limit the L/T formulation involves both the longitudinal and the transverse components, as the tensorial structures multiplying the coefficients $w (s, 0, s_2)$ are not independent as $s\to 0$. Obviously the final result, obtained with the correct limiting procedure, is the same in both cases.

Let's take in exam another kinematical configuration, more specific than the previous one, i.e. the case in which the two photons are both on-shell and massless or
\begin{itemize}
\item[b)] $s_1\, =\, s_2 \, = 0$ \qquad $s \neq 0$ \qquad $m =  0$.
\end {itemize}
In this case it is well known that the $AVV$ vertex exhibits a polar structure, as Dolgov and Zakharov showed in \cite{Dolgov:1971ri}, therefore we expect to recover this amplitude in the $s \to 0$ limit. The computed form factors are extremely simple. We obtain
\bea
A_1 (s, 0, 0) &=& - A_2 (s, 0, 0) = -\frac{i}{4 \pi^2},\\
A_3 (s, 0, 0) &=& - A_6 (s, 0, 0) = -\frac{i}{2 \pi^2 s}
\eea
which clearly exhibit the Bose symmetry for the two vector vertices, since $s_1=s_2$. Notice that $A_4$, $A_5$ are physically nonessential, as before; indeed they are multiplied, respectively, by $k_2^\nu$ and $k_1^\mu$  in the total amplitude $\Delta^{\la\mu\nu}(k_1,k_2)$, and vanish after their contraction with the physical polarization vectors of the photons.

The amplitude $\Delta^{\la\mu\nu}(k_1,k_2)$ satisfies the Ward identities written in Eq.~\ref{WI1}, since $s_{12} \to s/2 $ when both photons are on-shell
\bea
A_1 (s, 0, 0) = \frac{s}{2} \, A_3 (s, 0, 0), \qquad \qquad A_2 (s, 0, 0) = \frac{s}{2} \, A_6 (s, 0, 0).
\eea
In this case the entire correlator is obtained from only two form factors $A_i$ ($A_3$ and $A_6$), giving
\bea
\Delta^{\la \mu \nu} (s, 0, 0) &=& A_3 (s, 0, 0) \, \eta_3 ^{\la \mu \nu}  (k_1, k_2) + A_6 (s, 0, 0) \, \eta_6 ^{\la \mu \nu}  (k_1, k_2)  \nn \\
&=&  \frac{i}{2 \pi^2 s} \biggl[ \, k_2^{\mu} \veps[k_1, k_2, \nu, \la \, ]  \,  - \, k_1^{\nu} \veps[k_1,k_2, \mu, \la]\, \biggr] - \frac{i}{4 \pi^2}  \veps[(k_1-k_2), \la, \mu,\nu].
\eea
This expression can be reduced to its polar Dolgov-Zakharov form after using the Schouten identities in Eqs.~(\ref{schouten1},\ref{schouten2})
\bea
\Delta^{\la \mu \nu} (s, 0, 0)  = - \frac{i}{2 \pi^2 } \frac{ k^{\la}}{s} \, \veps[k_1, k_2, \mu, \nu]
\eea
as $s_1=s_2=0$.\\
In the L/T parameterization we expect a similar polar result,  after summing over the contributions coming both from the longitudinal and transverse tensors. In this case, the only two non-vanishing coefficients are $w_L$ and $w_T^{(+)}$
\bea
w_L(s, 0,0) &=& w_T^{(+)}(s, 0,0) = - \frac{4i}{s}, \\
w_T^{(-)}(s, 0,0) &=& \tilde{w}_T^{(-)}(s, 0,0) = 0
\eea
and the residues must be computed combining them with the corresponding tensor structures. It is worth noticing that  $t ^{(+)}_{\la \mu \nu} (k_1,k_2) =0$ for $s_1=s_2=0$. This can be immediately checked starting from its definition given in Eq. (\ref{calw}) and with the aid of the two Schouten identities shown in Eqs.(\ref{schouten1},\ref{schouten2}), which in this case become
\bea
k_1^ {\la}\, \veps[k_1,k_2,\mu, \nu]  &= &
- k_1^\nu \veps[k_1,k_2,\la, \mu]  \,
 + \frac{s}{2}  \veps[k_1,\la,\mu, \nu],   \\
k_2^ \la \, \veps[k_1,k_2,\mu, \nu]  &= &
k_2^\mu \,  \veps[k_1,k_2,\la, \nu]  - \,
\frac{s}{2} \veps [k_2,\la,\mu, \nu] ,
\eea
so that the unique contribution to the residue for $s\to0$ comes from the longitudinal part
\bea
\lim _{s \rightarrow 0 } \,  s \mathcal \, W _{\mu \nu \la} (s, 0, 0) &=&  \frac{1}{8 \pi^2}   \lim _{s \rightarrow 0 }  \,  s \, \,  \mathcal \, W^{L\, \lambda\mu\nu} \, \nn \\
&=&   \frac{1}{8 \pi^2} \lim_{s \rightarrow 0 } \,   s \, w_L(s, 0, 0) \, k_{\la} \veps[\mu, \nu, k_1, k_2] \nn \\
&=&  - \frac{i}{2 \pi^2 }  k^{\la} \, \veps[k_1, k_2, \mu, \nu].
\eea
We conclude that the pole is indeed present in the L/T amplitude if the conditions $s_1=s_2=0$ with $s\neq 0 $ are simultaneously satisfied
\bea
\Delta^{\la \mu \nu} (s, 0, 0)  = \mathcal \, W _{\mu \nu \la} (s, 0, 0)  =  - \frac{i}{2 \pi^2 } \frac{ k^{\la}}{s} \, \veps[k_1, k_2, \mu, \nu].
\eea

Another interesting case is represented by a symmetric kinematical configurations in which the external particles are massive gauge bosons of mass $M$. This will turn useful in the next sections, when we will discuss the behaviour of a BIM amplitude with massive external lines at high energy, showing, also in this case, its pole dominance. There are some conclusions that we can draw from this study which are important for the analysis of the next sections. Notice that in all the cases that we have discussed it is possible to isolate a $1/s$ contribution in
$w_L$ for any kinematical configurations other than the massless ($s\to 0$) one, where the L/T formulation requires a limiting procedure. This is clearly suggestive of the fact that a longitudinal component is intrinsically part of the vertex and not just of its collinear and chiral limit. This contributions is paralleled, in the Rosenberg amplitude(s)  by a constant behaviour of $A_1$ and $A_2$ ($A_1=i/(4 \pi^2)+...$). Massive external gauge lines or mass corrections due to the fermion mass in the loop do not shift this $1/s$ pole. 

As we have mentioned, under the general configurations contemplated in these last cases, these poles are not coupled in the IR, although this does not necessarily exclude a possible role played by these contributions in the IR region. However, the complete absence of a scale in their definition makes them suitable also of a completely different interpretation, as longitudinal contributions that survive in the asymptotic $s\to \infty$ limit of these amplitudes. In fact, we are going to show that any UV completion of these theories  has necessarily to deal with the cancellation of these terms.

\section{Pole dominated amplitudes: the UV significance of the general anomaly poles}

\begin{figure}[t]
\begin{center}
\includegraphics[width=15cm]{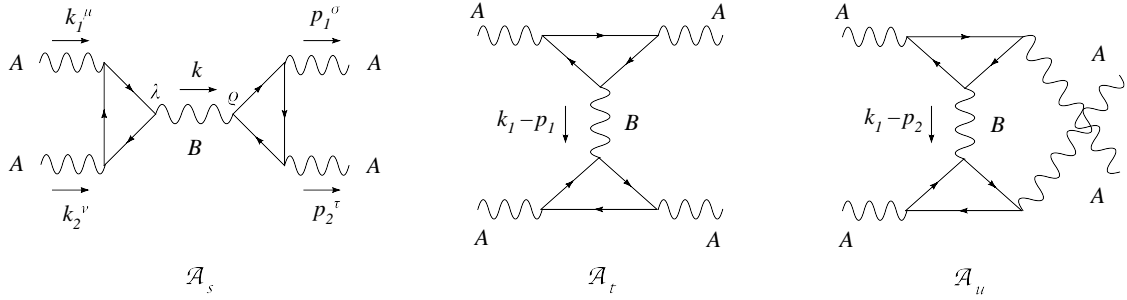}
\caption{\small The scattering process  $AA\rightarrow AA$ via a BIM amplitude in the three channels. The subscript $s, t, u$ stands for the channel. The exchanged gauge boson $B$ is different from the external ones and has a mass $M_B$.}
\label{AAAA}
\end{center}
\end{figure}

The UV significance of the poles appearing in the off-shell correlator can be established by studying a class of amplitudes that are pole-dominated at high energy, and which are typical of an anomalous theory (see Fig. \ref{AAAA}). These amplitudes describe the elastic scattering of massive (or massless) gauge bosons mediated by two triangle graphs and give total cross sections that grow quadratically with energy, thereby violating unitarity. 
As we have seen, for on-shell massless external gauge bosons (the $A$ lines) the anomaly vertex is characterized by a purely longitudinal component since the transverse form factors vanish. It is therefore obvious to conclude that $w_L$ is responsible for the high energy behaviour of these amplitudes. In this section we are going to show that a similar behaviour is found in the scattering of massive gauge bosons and that it can be attributed to the same component $w_L$ even though the transverse contributions, coming from the remaining form factors, are non vanishing. 

We start from the massless gauge bosons case and consider the BIM amplitudes  for the process $AA \rightarrow AA$ depicted in Fig.\ref{AAAA} for the three channels. The incoming momenta are $k_1^\mu$, $k_2^\nu$ in the initial state, while $p_1^\sigma$ and $p_2^\tau$ are those of the final state. The Mandelstam variables are defined as usual
\bea
s &=& (k_1+ k_2)^2 = (p_1+p_2)^2, \\
t &=& (k_1 - p_1)^2 = (k_2 - p_2)^2, \\
u &=& (k_1 - p_2)^2 = (k_2 - p_1)^2, \\
&& s + t + u = 0,
\eea

and we denote with $\theta$ the angle between the initial and final directions of the two particles in the center of mass frame.
 Each triangle reduces to its Dolgov-Zakharov form as the external lines are all massless and on-shell. Consider, for instance, the scattering mediated by a massive gauge boson $B$ in the $s$-channel, which is described by the amplitude
\bea
\mathcal A_s^{\mu\nu\sigma\tau} &=& \Delta^{\la\mu\nu}(-k,-k_1,-k_2)  \frac{1}{s - M_B^2}
 \biggl (g^{\la \rho} - \frac{k^\la k^{\rho}} {M_B^2} \biggr) \Delta^{\rho\sigma\tau}(k,p_1,p_2)
\eea
which becomes, after using the Ward identity on the axial-vector current
\bea
\mathcal  A_s^{\mu\nu\sigma\tau} &=&  \frac{a_n}{ M_B} \veps[\mu,\nu,k_1,k_2]  \frac{1}{s}  \frac{a_n} { M_B}\veps[\sigma, \tau,p_1,p_2].
\label{wwzz}
\eeqa
We perform a complete computation of the BIM amplitudes by combining all the $s$, $t$ and $u$ channels exchanges.
The amplitude with the exchange of $B$ in the three channels depicted in Fig.~\ref{AAAA} is given by
\bea
\mathcal M^{\mu \nu \sigma\tau}_{AA\rightarrow AA} &=&  (\mathcal A_s + \mathcal A_t + \mathcal A _u)^{\mu \nu \sigma\tau} ,
\eea
- where the subscript indicates the channel - and each term is composed by two triangle correlators and a Proca propagator of the exchanged $B$ gauge boson 
\bea
\mathcal A_s ^{\mu \nu \sigma\tau} &=&  \Delta ^{\mu \nu \la} \, (-k, -k_1,-k_2) \, P^{\la \rho } (k_1+k_2) \,
\Delta ^{\ro \si \tau} (k_1+k_2, p_1, p_2),
\label{As0} \\
 \mathcal A_t ^{\mu \nu \sigma\tau}&=& \Delta ^{\mu \sigma \la} (- (k_1-p_1), -k_1, p_1) \, P^{\la \rho } (k_1 - p_1) \,
 \Delta ^{\ro \tau \nu} (k_1 - p_1, p_2, -k_2), \\
 \label{At0}
\mathcal A_u ^{\mu \nu \sigma\tau}&=&  \Delta ^{\la \mu \tau} (- (k_1  -p_2), -k_1, p_2) \, P^{\la \rho } (k_1 - p_2) \,
\Delta ^{\ro \sigma \nu}(k_1 - p_2, p_1, -k_2).
 \label{Au0}
\eea
In the expressions above, the amplitude $\Delta$ is represented by a triangle correlator with external massless on-shell lines ($k_1^2 = k_2^2 =  p_1^2 = p_2^2 = 0$), which takes its polar (Dolgov-Zakharov) form
\bea
\Delta^{\la \mu \nu} (k, k_1, k_2)    =  a_n \frac{ k^{\la}}{s} \, \veps[k_1, k_2, \mu, \nu] ,
\qquad  \qquad  a_n=- \frac{i}{2\pi^2},
\label{DZ}
\eea
while the generic Proca propagator for the internal gauge boson $B$ with mass $M_B$ is
\bea
P^{\la \rho } (k) = - \frac{i}{k^2 - M_B^2} \, \biggl [g^{\la \rho} - \frac{k^\la k^{\rho}} {M_B^2} \biggr].
\label{prop}
\eea

After inserting the Eqs.~(\ref{DZ}) and (\ref{prop}) into Eqs.~(\ref{As0}-\ref{Au0}) we obtain for the single squared amplitudes and the interferences
\bea
&& |\mathcal A_s |^2 = 2 \, \hat a \, s^2  \qquad \qquad
|\mathcal A_t |^2 =  2 \, \hat a \, t^2  \qquad \qquad \,\,\,\,\,\,
|\mathcal A_u |^2 = 2 \, \hat a  \, (s+t)^2 \\
&& \mathcal A_s \mathcal A_t^* = \hat a \,  s \, t  \qquad \qquad
\mathcal A_s \mathcal A_u^* = - \hat a\,  s \, (s+ t)  \qquad
\mathcal A_t \mathcal A_u^* = - \hat a \, t \, (s+ t)
\eea
with $\hat a = {|a_n|^4}/{(8 M_B^4)} $, and then a  short computation yields
\bea
| \overline \mathcal  M |^2 _{AA \rightarrow AA} (s, \theta) &=&   \frac{1}{4} \sum _{spins} | \mathcal M ^{\mu \nu \sigma \tau}|^2  \nn \\ &=& \frac{|a_n|^4}{4 M_B^4} (s^2 + st + t^2) =
\frac{|a_n|^4}{64 } \frac{s^2}{M_B^4} (\cos^2 \theta + 3).
\label{mathcalM}
\eea
In Eq. \ref{mathcalM}) we have averaged over the initial states. The result depends on the total anomaly $a_n$ and on the St\"uckelberg mass of the exchanged gauge boson $M_B$; it takes the form
\bea
\frac{d \sigma}{d \Omega} = \frac{1}{2} \biggl(\frac{\hbar c }{8 \pi}\biggr)^2 \frac {| \overline \mathcal {M}|^2 (s, \theta) }{s},
\label{dsigma}
\eea
which violates the unitarity bound
\bea
\frac{d \sigma}{d \Omega} \leq \frac{1}{s}
\eea
as $s$ approaches infinity.
In an analogous way we deal with the case in which the gauge bosons $A$ are massive and satisfy on-shell conditions of the form $k_1^2 = k_2^2 =  p_1^2 = p_2^2 =  M^2$. The process is again the one depicted in Fig.\ref{AAAA} but the presence of massive external lines increases notably the length of the computation. The amplitude is neatly separated into longitudinal (polar) and transverse components. The longitudinal component is controlled by $w_L\sim 1/s$, which is multiplied by kinematical factors causing an overall growth of this component ($\sim s^2$) at large energy, while the transverse part behaves as
\bea
{w_T^{(+)}}(s)  \sim \frac{4 i}{s} \, \, \left( 1 + \log \frac{M^2} {s} \right) 
\eea
at large $s$. The transverse component of the squared amplitude has an overall $\sim 1/s^2$ behaviour in the same limit, and the corresponding amplitude can be correctly interpreted as due to the exchange of an ordinary massless propagator
($\sim1/s$). The threshold for this s-channel amplitude is at $s=4 M^2$, where it vanishes, while in the non-asymptotic region its transverse part describes the exchange of an ordinary $1/(s - M_B^2)$ propagator (times finite residues at each of the two vertices). In fact, the transverse component is well-behaved at any finite $s$ values and, in particular, for $s=M_B^2$. Notice also that in the limit $s\rightarrow 4 M^2$ (when $s>4 M^2$),
the function $\left|{w}_T^{(+)}(s)\right|^2$ does not exhibit poles and it can be written as
\ba
\left|{w}_T^{(+)}(s)\right|^2\sim \frac{a_1}{M^4} + a_2 \frac{a_2 s}{M^6} + \frac{a_3 s^2}{M^8} +\dots \,,
\ea
which implies the finiteness of the amplitude at threshold. As we have already mentioned, the same behaviour is
found at any finite value of $s$. Without enforcing the longitudinal subtraction, the cross section is unbound and the asymptotic expansion of the squared amplitude is
\bea
| \overline \mathcal  M |^2 _{AA \rightarrow AA} (s, \theta) \sim
\frac{1}{9}  \biggl[\frac{16} {M_B^4} (\cos \theta ^2+3) s^2\biggr],
\eea
where the term increasing linearly with $s$ (when inserted in the cross section) is dominated by the coefficient of  $|w_L |^2$. Therefore, the subtraction of the longitudinal component of the complete amplitude is necessary in order restore unitarity, leaving only the transverse part. The computations are rather lengthy, but the result for the transverse contributions, which respects unitarity at high energy, is given by the simple expression
\bea
{|\mathcal M |^2}_T &=&
\frac{M^4 \left(s-4 M^2\right)^2  \left(t^2+u^2\right) }  {2  \left(M_B ^2 - s\right)^2}\,\, |{w_T^{(+)}}(s)|^4
+\frac{M^4 \left(t-4  M^2\right)^2 \left(s^2+u^2\right) }{2   \left(M_B^2 - t\right)^2}\,\, |{w_T^{(+)}}(t)|^4 \nn \\
&+& \frac{M^4 \left(s^2+t^2\right) \left(u-4 M^2\right)^2  } {2 \left(M_B^2-u\right)^2} \,\, |{w_T^{(+)}}(u)|^4 \nn \\
&+& \frac{ M^4} {2 \left(M_B^2-s\right)  \left(M_B^2 - t \right)}
\left[ 128 M^8 - 64 (s+t) M^6+8 \left(s^2-3 t s+t^2\right) M^4  \right. \nn \\
&&\hspace{2.5cm }  + \left. 6 s t (s+t) M^2 +s   t \left(s^2+3 t s+t^2 \right) \right]\, \, |{w_T^{(+)}}(s)|^2   \, \, |{w_T^{(+)}}(t)| ^2 \nn \\
&+& \frac{ M^4}  {2 \left(M_B^2  -   s\right)  \left(M_B^2 - u \right)}
\left[128 M^8 - 64 (s+u) M^6 + 8 \left(s^2-3 u s+u^2\right)   M^4 \right. \nn \\
&& \hspace {2.5cm} + \left. 6 s u (s+u) M^2+s u \left(s^2+3 u   s+u^2\right)\right] \, \, |{w_T^{(+)}}(s)|^2 \,\,| {w_T^{(+)}}(u)|^2 \nn \\
&+&\frac{ M^4}  {2 \left(M_B^2 - t   \right)  \left(M_B^2 - u \right)}
\left[128 M^8 - 64 (t+u) M^6+8 \left(t^2-3 u t+u^2\right)   M^4 \right. \nn \\
&& \hspace{2.5cm} + \left. 6 t u (t+u) M^2+t u \left(t^2+3 u  t+u^2\right)\right] \, \, |{w_T^{(+)}}(t)|^2 \,\, |{w_T^{(+)}}(u)|^2. \nn \\
\label{Mtrans}
\eea

Notice that the leading terms for ${w_T^{(+)}}(t) $ and ${w_T^{(+)}}(u) $ in the asymptotic region are the same as those contained in ${w_T^{(+)}}(s) $. Expressing in terms of $s$ and the scattering angle in the center of mass frame $\cos \theta$ all the other invariants
\bea
t = \biggl [2 M^2 - \frac{s}{2} \biggr ] (1 - \cos \theta) \qquad \qquad u =  \biggl [2 M^2 - \frac{s}{2}\biggr ] (1 + \cos \theta);
\eea
 Eq.~\ref{Mtrans} shows that ${|\mathcal M |^2}_T  \rightarrow 0$  for $s \rightarrow \infty$,
which is in agreement with unitarity. At the same time, the interpretation of the corresponding squared amplitude in terms of an ordinary bosonic exchange is rather obvious since the purely transverse part shows an asymptotic behaviour of the form
\ba
&&\left|{\cal M}\right|^2_T\sim \frac{M^4}{s^2}\sum_{n=0} C_n(\theta,M) \log^n\left(\frac{M^2}{s}\right)
\ea
with the correctly factorized double pole ($\sim 1/s^2$), and where the coefficients $c_n(\theta,M)$ depend only on the mass $M$ of the external lines
and on the scattering angle.

To summarize, we have seen that anomaly poles extracted by a complete off-shell analysis of the correlation function of the anomaly graph have a clear UV significance and saturate the anomaly. We conclude that the anomaly diagram can always be written in terms of an anomaly pole plus extra terms which either contribute homogeneously to the anomalous Ward identity or are responsible for its mass corrections. These two sources of breaking of the Ward identity have separate origin and appear to be universal. We are going to use this result to present a form of the effective action which includes these extra contributions, by discussing several of its expansions using as an expansion parameter the mass of the fermion in the loop.

\section{Effective actions and the gauge anomaly}
In this section we are going to discuss the formulation of the effective action in the presence of anomaly poles, generalizing the Euler-Heisenberg (EH) result to an anomalous theory. We will focus our attention exclusively on the 
trilinear gauge terms, coming from the anomalous structure, which are new compared to the EH formulation. 
 
The simplest example that we can consider is a theory describing a single anomalous gauge boson $B$ with a lagrangian
\beq
\mathcal{L}_{B}= \overline{\psi} \left( i \, \slash{\partial} + e \slash{B} \gamma_5\right)\psi - \frac{1}{4} F_B^2.
\label{count0}
\eeq
The effective action of the model suffers from a trilinear gauge interaction which is anomalous ($BBB$). In this case the anomalous vertex is obtained by
a simple symmetrization of (\ref{Ros}) which generates a $\Delta_{AAA}$ vertex
\beq
\Delta_{AAA}= \frac{1}{3}\left(\Delta_{AVV} + \Delta_{VAV} + \Delta_{VVA}\right).
\eeq

The anomalous gauge variation $(\delta B_\mu=\partial_\mu\theta_B)$
\beq
\delta\Gamma_B = \frac{ i \, e^3 \, a_n}{24} \, \int d^4 x \, \theta_B(x) \, F_B\wedge F_B
\label{var1}
\eeq
can be reproduced by the nonlocal action
\beq
{\Gamma}_{pole}= \frac{e^3}{48 \, \pi^2} \, \langle \partial B(x) \square^{-1}(x-y) F_B(y)\wedge F_B(y)
\rangle,
\label{var2}
\eeq
 which is the variational solution of (\ref{var1}). To derive a $1/m$ expansion of the effective action, we perform an expansion of the Rosenberg form factors, obtaining 
\bea
A_1 (s,0,0,m^2) &=& - A_2 (s,0,0,m^2) = \frac{i }{48 \pi ^2} \frac{s}{m^2} +\frac{i }{360 \pi ^2} \frac{s^2}{m^4} +O\left(\frac{1}{m^6}\right), \label{expA1}\\
A_3 (s,0,0,m^2) &=& - A_6 (s,0,0,m^2) = \frac{i }{24 \pi ^2}\frac{1}{m^2} +\frac{i }{180 \pi ^2}\frac{s}{m^4}+ O\left(\frac{1}{m^6}\right), 
\\
A_4 (s,0,0,m^2) &=& - A_5 (s,0,0,m^2) = \frac{i }{12 \pi ^2}\frac{1}{m^2} +\frac{i}{120 \pi ^2}\frac{s}{m^4} +O\left(\frac{1}{m^6}\right),
\label{expA4}
\eea
where $s \equiv  k^2$. We will also use the notation $s_1$ and $s_2$ to denote the virtuality of the two external photons ($s_1 \equiv k_1^2,  s_2 \equiv k_2^2$). Due to the chiral gauge anomaly, the effective action is gauge-variant. For our choice of momenta
(incoming $k$ on the axial-vector of index $\lambda$ and outgoing $k_1$ and $k_2$ on
 the two vector currents of indices $\mu$ and $\nu$) we obtain
\bea
T^{\lambda\mu\nu}_{AVV}(x,y,z) =  \int \frac {d^4 k \, d^4 k_1 \, d^4 k_2} {(2 \pi)^8} \, \delta^4 (k-k_1-k_2) \, e^{i k \cdot z - i k_1 \cdot x - i k_2 \cdot y} \, \Delta^{\la \mu\nu}_{AVV} (k,k_1,k_2)
\eea
with the contribution of the anomalous vertex being given by
\beq
\Gamma^{(3)}= - \frac{i}{6} \, \int d^4 x \,d^4 y \, d^4 z \, T^{\lambda\mu\nu}(x,y,z) B_\lambda(z) \, B_\mu(x) \, B_\nu (y),
\eeq
where $T^{\lambda\mu\nu}(x,y,z) $ is the symmetrized correlator given by
\beq
T^{\lambda\mu\nu}(x,y,z)  = \frac{1}{3} \biggl[ T_{AVV}^{\lambda\mu\nu}(x,y,z) + T^{\lambda\mu\nu}_{VAV} (x,y,z)  + T^{\lambda\mu\nu}_{VVA} (x,y,z) \biggr]. 
\eeq
The explicit form of the new anomalous contributions (the symbols $\langle\,\, \rangle$ denote spacetime integration) can be obtained by plugging in the expression of the various form factors expanded in $1/m$ written in Eqs.(\ref{expA1}-\ref{expA4}). We obtain
 \beqa
 \Gamma^{(3)} &=&- \frac{i}{6} \biggl[ \frac{1}{48 \pi^2 m^2}\epsilon^{\alpha\mu\nu\lambda}\left(
 \langle \square B_\lambda \partial_\alpha B_\mu B_\nu \rangle -
 \langle \square B_\lambda B_\mu \partial_\alpha B_\nu \rangle \right) \nonumber \\
&& -\frac{1}{360 \pi^2 m^4}\epsilon^{\alpha\mu\nu\lambda} \left( \langle \square^2 B_\lambda \partial_\alpha B_\mu B_\nu\rangle -
\langle \square^2 B_\lambda B_\mu\partial_\alpha B_\nu\rangle \right)\nonumber \\
&& + \frac{1}{24 \pi^2 m^2}\left(\epsilon^{\alpha\beta\mu\lambda}\langle \partial_\alpha\partial_\nu B_\mu B_\lambda
\partial_\beta B^\nu\rangle - \epsilon^{\alpha\beta\nu\lambda}\langle \partial_\alpha B_\mu B_\lambda \partial_\beta \partial^\mu B_\nu\rangle\right) \nonumber \\
&& -\frac{1}{180 \pi^2 m^4}\left( \epsilon^{\alpha\beta \mu\lambda}\langle \partial_\alpha\partial_\nu B_\mu \square B_\lambda \partial_\beta B^\nu\rangle - \epsilon^{\alpha\beta \nu\lambda}\langle \partial_\alpha B_\mu \square B_\lambda \partial_\beta \partial^\mu B_\nu \rangle \right) \nonumber \\
&& +\frac{1}{12 \pi^2 m^2}\left(
\epsilon^{\alpha\beta\mu\lambda}\langle \partial_\alpha B_\mu \partial_\beta \partial_\nu B^\nu B_\lambda\rangle
-\epsilon^{\alpha\beta\nu\lambda} \langle\partial_\alpha \partial_\mu B^\mu B_\lambda \partial_\beta B_\nu\rangle
\right) \nonumber \\
&& -\frac{1}{120 \pi^2 m^4} \left(\epsilon^{\alpha \beta\mu\lambda}\langle \partial_\alpha B_\mu \partial_\beta\partial_\nu \square B_\lambda\rangle - \epsilon^{\alpha\beta\nu\lambda} \langle
\partial_\alpha \partial_\mu B^\mu \square B_\lambda \partial_\beta B_\nu\rangle \right)\biggr].
\eeqa 
  
Naturally, the $p/m$ expansion hides the nonlocal contributions which are present in the effective action. These can be identified from the off-shell expression of the anomaly vertex, which in the L/T parameterization takes a close form only in momentum space. For this reason we rewrite this parameterization as a pole ($w_L=-4 i/s$) plus mass corrections in the equivalent form
\beq
 \mathcal \, W^{L\, \lambda\mu\nu}= \left(w_L  - \mathcal{F}(k, k_1, k_2, m)\right)  \, k^\lambda \veps[\mu,\nu,k_1,k_2]
\eeq
\beq
\mathcal{F}(m, s,s_1,s_2)= \frac{8 m^2}{\pi ^2 s} C_0(s, s_1,s_2,m^2),
\eeq
where $ C_0$ has been given in Eq. (\ref{C0polylog}). Obviously, the anomaly is completely given by $w_L$. The complete action is instead given by
\beq
 \Gamma^{(3)}=  \Gamma^{(3)}_{pole} + \tilde{\Gamma}^{(3)}
 \eeq
 with the pole part given by
\beq
\Gamma^{(3)}_{pole}= -\frac{1}{8 \pi^2} \int d^4 x \, d^4 y  \,\partial \cdot B(x) \square^{-1}_{x,y} F(y) \wedge F(y)
\label{gammapole}
\eeq
and the rest ($\tilde{\Gamma}^{(3)}$) given by a complicated nonlocal expression which contributes homogeneously to the Ward identify of the anomaly graph
 \beqa
  \tilde{\Gamma}^{(3)}&=&- \frac{e^3}{48 \pi^2 }\int d^4 x \, d^4 y \, d^4 z\, \partial \cdot B(z) F_B(x)\wedge F_B (y)
  \int \frac{d^4 k_1 \, d^4 k_2 }{(2 \pi)^8} \, e^{-i k_1 \cdot (x-z) - i k_2\cdot (y-z)} \mathcal{F}(k,k_1,k_2, m)
\nonumber \\
&&  - \frac{e^3}{48 \pi^2 } \int d^4 x \, d^4 y \, d^4 z  B_\lambda(z) B_\mu(x) B_\nu(y)
\int \frac{d^4 k_1 \, d^4 k_2}{(2 \pi)^8} \, \, e^{-i k_1 \cdot (x-z) - i k_2\cdot (y-z)} W_T^{\lambda\mu\nu}(k,k_1,k_2,m),\nonumber \\\label{gammafull}
 \eeqa
 where $k=k_1 + k_2$.  
 A second form of the effective action is obtained by expanding around $m=0$, i.e. for a small mass. A simple, but very instructive case,  is the one with two on-shell photons ($s_1=s_2=0$) and a nonzero fermion mass. We obtain, for instance, in the $AVV$ case the following expressions for the form factors after the series expansion around $m=0$ 
\bea
w_L &=& - \frac{ 4 \, i }{s} - \frac{ 4 \, i \, m^2}{s^2} \log \left( - \frac{s}{m^2}\right) + O (m^3), \\
w_T ^{(+)} (s,0,0,m^2 ) &=& \frac{12 \, i }{s} \, - \frac{4 \, i}{s}  \, \log \left( - \frac{s}{m^2}\right) + \frac{ 4 \, i \, m^2}{s^2} \left[ 2 +  \log \left(\frac{s^2}{m^4}\right) -  \log^2 \left( - \frac{s}{m^2}\right) \right] + O (m^3). \nn \\
\eea
It is clear that this second expansion allows to isolate the pole term from the mass corrections, and is probably a more faithful description of the anomalous content of the theory, identified by the anomaly pole. 

\section{Anomaly inflow from 5-D and the breaking of unitarity in the effective action }
The presence of a longitudinal exchange in an anomalous theory - which exhibits a power-like growth with energy of some of its S-matrix elements -   is not a property just of four dimensional models. As we are going to show, similar features are typical also of extra dimensional models in which the presence of anomalies on the branes, due to the delocalization of the chiral fermions, is canceled by an anomaly inflow. In particular, the presence of anomaly poles in the reduced theory is, in general, a threat to the consistency of the effective action. For instance, in 5-D models, the basic role of the mechanism of inflow is to guarantee the gauge invariance of the effective 4-D geometric action (after compactification), canceling the anomaly of the chiral fermions on the branes.
Our analysis, to be definite, is focused on a model in 5-D which shows a nice realization of the inflow, formulated in \cite{Hill:2006ei}, although our conclusions are expected to be model independent.  We are going to show that in the case of anomalous models with an inflow, any effective theory defined by a restriction on the sum over the KK modes is necessarily going to break unitarity in the UV because of the presence of pole dominated amplitudes, quite similarly to our previous analysis in 4-D. 

In general, it is well known that models incorporating extra dimensions violate unitarity both before and after compactification \cite{SekharChivukula:2001hz,Chivukula:2003kq, Muck:2001yv}. However, this stronger form of breaking obtained {\em for any fixed number} of KK modes included in the expansion, which 
does not occur for other (non anomalous) models of this type, finds its origin in the limitation of the condition of gauge invariance, here guaranteed by an inflow, to establish the full consistency of the theory.  The point that we will be raising is that an inflow has necessarily to remove the anomaly poles of the effective theory on the brane in order to make it a consistent model. Notice that this breaking of unitarity that we will discover is completely unrelated to other unitarity bounds that the theory obviously has for being non-renormalizable.
We follow closely \cite{Hill:2006ei}, skipping details that can be found in that work,  and consider the lagrangian
\begin{equation}
\label{freelagrangian}
\mathcal{L}(x, y) \, = \, - \frac{1}{4 \tilde{e}^2} F_{M N} (x, y) F^{M N} (x, y),
\end{equation}
where
\begin{equation}
\label{fieldstrength}
F_{M N} (x, y) \, = \, \partial_M A_N (x, y) - \partial_N A_M (x, y)
\end{equation}
denotes the 5-D field strength,
 Lorentz indices in 5-D are denoted with
capital Roman  letters, e.g.~$M,N  = 0,1,2,3,5$, while  the corresponding greek indices are four dimensional ($\mu,\nu  =  0,1,2,3$).  We use the notation
 $x  \equiv (x^0,\vec{x})$  and  $y \equiv x^5$  to  denote  the
coordinates  of the  usual $(3+ 1)$-dimensional spacetime  and the
coordinate of the orbifold, respectively. 
The nonzero KK modes acquire a typical St\"uckelberg mass due to the compactification.
One of the possible ways to realize an inflow in this model for the restoration of gauge invariance is by the introduction in 5-D of a Chern-Simons form (CS)
\bea
L_{CS} & = & \frac{\kappa}{4} \veps^{ABCDE} A_{A}F_{BC} F_{DE}.
\eea

Integrating over the $y$ dimension we obtain the effective 4-dimensional lagrangian
\bea
\mathcal{L} (x) &=&
\bigl [ \;\bar{\psi}
(i\slash{\partial}+ \slash{V} +
\slash{{\cal{A}}}\gamma^5-m)\psi
+ \frac{1}{12\pi^2} \sum_{nmk}c_{nmk}B^n_\mu B^m_\nu
\widetilde{F}^{k\mu\nu}
\nonumber \\ & &
-\frac{1}{4e^2} F^0_{\mu\nu}F^{0\mu\nu}
-\frac{1}{4e'{}^2}\sum_{n\geq 1} F^n_{\mu\nu}F^{n\mu\nu}
+\sum_{n\geq 1} \frac{1}{2e_n^2}M_n^2 B_\mu^n B^{n\mu}  \bigr ]
\eea
which describes a massless photon, whose field-strength is denoted by $F^0$, plus the corresponding Kaluza-Klein (KK) excitations ($F^n$) - which are massive - and the
infinite set of 4-D Chern-Simons terms. These are characterized by some numerical coefficients $c_{nmk}$ whose expression can be found in \cite{Hill:2006ei}. It is easily found that the 1-loop effective action of the model contains the infinite set of diagrams
\beq
T_{l,m,n}= \langle J_{A}^{(l)}J_A^{(m)}J_A^{(n)}\rangle,
\eeq
where the currents include, besides the vector ($J^{(n)}$) and axial-vector contributions ($J_A^{(n)}$) coupled to the KK gauge excitations, also the Chern-Simons part. As discussed in \cite{Hill:2006ei}, by absorbing a Chern-Simons term in the current (which amounts to induce some shifts in the $A_1$ and $A_2$ coefficients of Rosenberg, see also the discussion in \cite{Armillis:2007tb}) we can always bring the vertex correlator, also in this more general case, to reproduce Bardeen's result for the axial vector anomaly, moving all the anomaly of the vertex on the axial part. For this reason, in the analysis presented below, we will omit any explicit Chern-Simons term, having these been absorbed into the definition of the anomaly vertices - expressed in terms of vector and axial-vector currents rather than of chiral currents -  with conserved vector currents.

\begin{figure}[t]
\begin{center}
\includegraphics[scale=1.0]{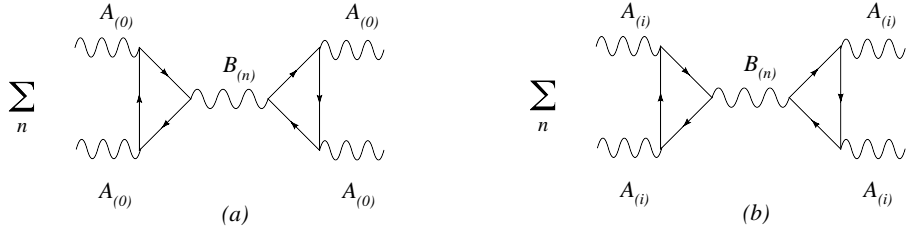}
\caption{\small BIM amplitude in the presence of a KK tower of modes exchanged in the s-channel. In a) the external zero modes $A_{(0)}$ are massless while in b) they have a fixed even KK parity $(i)$  and vector couplings with the fermions in the loop.}
\label{BIMKK}
\end{center}
\end{figure}

The breaking of unitarity can be established rigorously by considering a scattering amplitude which is pole dominated (due to the anomalies localized on the branes), where we now allow in the various channel the possibility of exchanging a finite number of KK excitations ($N_{KK}$). For this purpose we define two functions $f_N$ and $\chi_N$ defined as
\bea
\chi_{N_{KK}} &=& \sum_{odd \, n}^{N_{KK}} \frac{1}{M_n^2} \, , \\
f_{N_{KK}}(s) &=& \sum_{odd \, n}^{N_{KK}} \frac{1}{s - M_n^2}.  
\eea 
We can consider scatterings involving both massless and massive external gauge bosons in 2-to-2 amplitudes, illustrated in 
Fig.~\ref{BIMKK} for the s-channel case. For simplicity we present below only the computation in the massive case, although similar conclusions can be reached also in the massless one. We obtain

\bea
|\mathcal A_s |^2 &=&
\frac{1}{4} \, \left(s - 4 M^2\right)^2
\left[  s^4 \chi_N^2  |{w_L}(s)|^4 + 2 M^4 \left(t^2 + u^2 \right) f_N(s)^2 \, |{w_T^{(+)}}(s)|^4 \right], \\
|\mathcal A_t |^2 &=&
\frac{1}{4} \, \left(t - 4 M^2\right)^2
\left[  t^4 \chi_N^2  |{w_L}(t)|^4 + 2 M^4 \left(s^2 + u^2 \right) f_N(t)^2 \, |{w_T^{(+)}}(t)|^4 \right], \\
|\mathcal A_u |^2  &=&
\frac{1}{4} \, \left(u - 4 M^2\right)^2
\left[  u^4 \chi_N^2  |{w_L}(u)|^4 + 2 M^4 \left(s^2 + t^2 \right) f_N(u)^2 \, |{w_T^{(+)}}(u)|^4 \right], \\
\mathcal A_s \mathcal A_t ^*  &=&
-\frac{1}{2} M^4 u  \, t^2 \left (u - t \right) \chi_N f_N(s) \, \, | {w_L}(t)|^2\, \, |{w_T^{(+)}}(s)|^2 \nn \\
&& -\frac{1}{2} M^4 u  \, s^2 \left (u - s \right) \chi_N f_N(t)  \, \,|{w_L}(s)|^2 \, \, |{w_T^{(+)}}(t)|^2 \nn \\
&& + \frac{1}{4} M^4 f_N(s) f_N(t) \,
 \biggl[8 \left( 2 s^2-7 u s+u^2 \right) M^4 + 2 s \left( -4 s^2+3 u s+9 u^2 \right) M^2 \biggr. \nn \\
 && \biggl. \hspace{4cm} + s^4 - s u^3 + 2 s^3 u \biggr]  \, \,
   |{w_T^{(+)}}(s)|^2  \, \, |{w_T^{(+)}}(t)|^2 \, \nn \\
    && + \frac{1}{8} s^3 t^3 \chi_N^2 \, \, |{w_L}(s)|^2\,\,  |{w_L}(t)|^2,   \nn \\ \\
\mathcal A_s \mathcal A_u ^* &=&
\frac{1}{2} M^4 t \, u^2 \left (u - t \right) \chi_N f_N(s)  \, \, | {w_L}(u)|^2\, \, |{w_T^{(+)}}(s)|^2 \nn \\
&& +\frac{1}{2} M^4 t \, s^2 \left( s-t \right) \chi_N f_N(u) \, \,|{w_L}(s)|^2 \, \, |{w_T^{(+)}}(u)|^2 \nn \\
&& + \frac{1}{4} M^4 f_N(s) f_N(u) \,  \,
\biggl[8 \left(2 s^2-7 s t+t^2\right) M^4+2 s \left(-4 s^2+3 s t+9 t^2\right) M^2  \biggr. \nn \\
&& \biggl. \hspace{4cm} + s^4 - s t^3 + 2 s^3 t \biggr] \, \,
    |{w_T^{(+)}}(s)|^2  \, \, |{w_T^{(+)}}(u)|^2 \, \nn \\
 &&  + \frac{1}{8} s^3 u^3 \chi_N^2 \, \, |{w_L}(s)|^2\,\,  |{w_L}(u)|^2,   \nn \\ \\
\mathcal A_t \mathcal A_u ^* &=&
\frac{1}{2} M^4 s \, u^2 \left (u - s \right) \chi_N f_N(t) \, \, | {w_L}(u)|^2\, \, |{w_T^{(+)}}(t)|^2 \nn \\
&& +\frac{1}{2} M^4 s \, t^2 \left( t-s \right) \chi_N f_N(u) \, \,|{w_L}(t)|^2 \, \, |{w_T^{(+)}}(u)|^2 \nn \\
&& + \frac{1}{4} M^4 \, f_N(t) f_N(u)   \,
\biggl[8 \left(2 t^2-7 t s+s^2 \right) M^4 + 2 t \left(-4 t^2+3 t s+9 s^2\right) M^2 \biggr.\nn \\
&& \biggl. \hspace{4cm} + t^4 - s^3 t+2 s t^3 \biggr] \, \,
   |{w_T^{(+)}}(t)|^2 \,\, |{w_T^{(+)}}(u)|^2 \, \nn \\
   && + \frac{1}{8} t^3 u^3 \chi_N^2 \, \, |{w_L}(t)|^2\,\,  |{w_L}(u)|^2.   \nn \\
\eeqa 
Notice that $f_{N_{KK}}$ is well-behaved at large energy, due to the presence of a partial sum. Using the results of the previous section on the asymptotic behaviour of $w_T$, it is easy to figure out that {\em by removing this component} the cross section constructed by summing over all the squared amplitudes given above respects unitarity in the UV. The conclusions of this analysis are rather obvious: the appearance of anomaly poles in extra dimensional models, even for a gauge invariant lagrangian whose anomaly on the brane has been cured by the inflow,  implies the presence of additional unitarity bounds on the effective actions of the induced (4-dimensional) theory on the brane. Therefore its completion requires necessarily the entire extra-dimensional construct. As we have already remarked above, the breaking of unitarity induced by the presence of these amplitudes (and poles) is unrelated to other sources of breaking, attributed in the past to the sum over the KK excitations.

\section{Conclusions}
The presence of anomaly poles in the perturbative expansion of the effective action, appears to be an essential property of anomalous theories, even in the most general kinematical configurations of the anomalous correlators. We have shown that only a complete computation of the effective action allows to identify such contributions, which affect the UV behaviour of a correlator even if they are decoupled in the IR. The goal of our work has been to show that more general anomaly poles are present in the perturbative description of the anomaly.
Previously, the appearance of these terms was considered a pure IR phenomenon, while their isolation in the L/T parameterization was probably considered an artificial result due to the presence of Schouten relations in the anomaly graph. We have also shown how the Schouten relations can 
"dissolve" a pole, by allowing its rewriting in terms of additional form factors which are not of polar form. 
However, we have shown that the true meaning of the pole and its irreducibility becomes evident from the UV study of some amplitudes. These are pole dominated and become harmless at high energy only by the subtraction of the longitudinal components induced by the presence of these terms.  A similar result holds also for models with extra dimensions when a mechanism of inflow is invoked to restore gauge invariance of an anomalous theory on the brane. 

In our work we have performed a complete and very detailed analysis of all the relevant regions of the anomaly graph, identifying all the relevant sources of singularities in the correlator and generalized the L/T parameterization to the massive case. This result has been used to derive an effective action which generalizes the Euler-Heisenberg result to anomalous theories. In a companion work we are going to investigate the significance of anomaly poles in the case of conformal anomaly, showing the perfect (and striking) analogy with the patterns of anomaly poles discussed in this work. 

\centerline{\bf Acknowledgements}
We thank  Nikos Irges, Fiorenzo Bastianelli, Alan White and Theodore Tomaras  for discussions and Chris Hill for some clarifying comments related to his analysis. C.C. thanks Roberto Soldati and Alexander Andrianov for discussions during a visit at the University of Bologna.
This work was supported (in part) by the EU grants INTERREG IIIA (Greece-Cyprus) and by the European Union through the Marie Curie Research and Training Network ``Universenet'' (MRTN-CT-2006-035863). C.C. thanks the University of Crete for hospitality.

\section{Appendix A. Poles and residui for massive gauge bosons}
\label{poles_residui}
We are interested in the limit
\begin{itemize}
\item[c)] $s_1 \, = \, s_2\, = \, M^2$ \qquad $s  \neq 0$ \qquad $m =  0$.
\end {itemize}
In this case only few simplifications occur in the complete expressions of the amplitudes $A_i$
since the only surviving symmetry is the one between $s_1$ and $s_2$ and  no momentum
is set to zero. The expansion of the three point function is the most general one and the invariant amplitudes are given by
\bea
A_1 (s, M^2, M^2) &=& -\frac{i}{4 \pi^2}\\
A_3 (s, M^2, M^2) &=&
- \frac{2\,  i \, M^4\,   } {\pi ^2 s^2 \left(s-4 M^2\right)^2} \, \,  \Phi_M   \, \, (s - M^2 ) \nn \\
&& -  \frac{i}   {2 \pi ^2 s \left(s-4 M^2\right)^2}
 \left[ s^2  - 6 s M^2+2  \left(2 M^2+s\right) \log  \left[\frac{M^2}{s}\right] M^2+8 M^4\right] \nn \\ \\
A_4 (s, M^2, M^2) &=&
\frac{i  M^2 }  {\pi ^2 s^2 \left(s - 4 M^2\right)^2}
\Phi_M \left(s^2 -3 s M^2 + 2 M^4 \right)  \nn \\
&& +\frac  {i}   {2 \pi ^2 s \left(s-4 M^2\right)^2}
\left[  2 s M^2  + \left(s^2-4 M^4\right) \log \left(\frac{M^2}{s} \right) -8 M^4  \right],
\eea
with the functions $\Phi(x,y)$ and $\lambda(x,y)$ defined in this  specific case by
\bea
 \Phi_M \equiv \Phi (  \frac{M^2}{s} , \frac{M^2}{s}) &=& \frac{1}{\lambda_M}
\left[ \log ^2 \left(\frac{2 M^2} {s ( \lambda_M +1)  -2 M^2} \right) + 4 \textrm{Li}_2\left(\frac{2 M^2}
{- s ( \lambda_M +1) + 2 M^2 }\right)+ \frac{\pi ^2}{3}  \right], \nn \\ \\
\lambda_M &\equiv& \lambda(M^2/s, M^2/s) = \sqrt{1 - \frac{4 M^2}{s }},
\eea
as in Eqs.~(\ref{Phi},\ref{lambda}), with $x=y=M^2/s$.

As usual, a symmetric configuration of this type yields
\bea
A_2 (s, M^2, M^2) &=& - A_1 (s, M^2, M^2), \\
A_5 (s, M^2, M^2) &=& - A_4 (s, M^2, M^2),  \\
A_6 (s, M^2, M^2) &=& - A_3 (s, M^2, M^2)
\eea
and in the total amplitude only few simplifications occur
\bea
\Delta^{\la \mu \nu} (s, M^2, M^2) = A_3 (s, M^2, M^2) \, \eta_3 ^{\la \mu \nu}  (k_1, k_2) +
A_4 (s, M^2, M^2) \, \eta_4 ^{\la \mu \nu}  (k_1, k_2)  \nn \\
+ A_5 (s, M^2, M^2) \, \eta_5 ^{\la \mu \nu}  (k_1, k_2)
+ A_6 (s, M^2, M^2) \, \eta_6 ^{\la \mu \nu}  (k_1, k_2).
\eea
The analysis of the spurious pole at $s=0$ requires the analytic continuation in the euclidean region ($s<0$) according to the $i \eta$ prescription: $ s \to s+ i \eta $, $M^2 \to M^2 + i \eta$. In this case the only trascendental functions requiring the analytic regularizations are the logarithmic ones, the dilogarithm being well-definite since
\bea
\frac{2 M^2} {- s ( \lambda_M +1) + 2 M^2 }  < 1 \qquad \textrm{for}\, s<0.
\eea
Then we substitute
\bea
\log \left [\frac{M^2}{s} - i \eta	\right ] &\to&  \log\left[- \frac{M^2}{s} \right] - i \pi  \qquad \textrm{for}\, s<0
\label{anlog} \\
 \log \left[\frac{2 M^2} {-2 M^2+s+s \lambda }  - i \eta \right] &\to&  \log \left[- \frac{2 M^2} {-2 M^2+s+s \lambda }\right]  - i \pi   \qquad \textrm{for}\, s<0
 \label{anlog2}
\eea
into the expressions of  $A_3 (s, M^2, M^2)$ and  $A_4 (s, M^2, M^2)$ and perform the limit for $s \to 0$. We obtain
\bea
&& \lim_{s \rightarrow 0 } \,   s \, A_i (s, M^2, M^2)
= 0 \qquad \qquad i = 3, \dots, 6
\eea
and also
\bea
\lim_{s \rightarrow 0 } \,   s \, \Delta^{\la \mu \nu} (s, M^2, M^2) = 0,
\eea
showing that in the presence of external massive gauge lines the triangle amplitude $\Delta^{\la \mu \nu}$ exhibits no poles.
This can be confirmed by a parallel analysis based on the L/T parameterization whose coefficients are
\bea
w_L(s, M^2,M^2) &=& - \frac{4i}{s}, \\
w_T^{(+)}(s, M^2,M^2) &=& \frac{4 i}{(s-4M^2)^2}\left[(s+2M^2)\log\left[\frac{M^2}{s}\right]+\frac{2M^2(s-M^2)}{s}\Phi_M \right]\nn \\
&+& \frac{4 i}{s-4M^2},\\
w_T^{(-)}(s, M^2,M^2) &=& \tilde{w}_T^{(-)}(s, M^2,M^2) = 0.
\eea
Combining the previous results, the whole amplitude becomes
\beq
 \mathcal \, W ^{\lambda\mu\nu} \, (s, M^2,M^2)  =
  \frac{1}{8\pi^2}\left [   w_L (s, M^2,M^2) \, k^\lambda \veps[\mu,\nu,k_1,k_2] - w_T^{(+)}(s, M^2,M^2) \,t^{(+)}_{\lambda\mu\nu}(k_1,k_2) \right].
\eeq
At this point we perform the same analytic continuations discussed above, shown in Eqs.~(\ref{anlog},\ref{anlog2}) and take the limits
 \bea
&& \lim_{s \rightarrow 0 } \,  s \,  w_L (s, M^2, M^2) = - \, 4 \, i \\
&& \lim_{s \rightarrow 0 } \,   s \,  w_T^{(+)} \, (s, M^2, M^22) \,  t _{\la \mu \nu} ^{(+)} \, (k_1, k_2) =   - \, 4 \, i
\eea
which, in combination, give a vanishing residue also in this parameterization
\bea
 \lim_{s \rightarrow 0 } \,  s \, \mathcal \, W ^{\lambda\mu\nu} \, (s, M^2,M^2) = 0.
\eea
\\
When the mass of the fermion in the loop is non vanishing, $m\neq0$,  we consider cases $d)$, $e)$ and $f)$. We take the appropriate limits starting from the expressions in Eq.(\ref{A1mass}-\ref{A5mass}) obtaining
\begin{itemize}
\item [d)] $k_1^2 =0$ \qquad $k_2^2 \neq 0$ \qquad $k^2 \neq 0$ \qquad $m\neq 0$
\end{itemize}
\bea
A_1 (s, 0, s_2, m^2) &=& -\frac{i}{4 \pi ^2} +\frac{s_2}{4 \pi ^4 \left(s-s_2\right)} \, D_2  -\frac{m^2}{2 \pi^4} \, \bar C_0, \\
A_2 (s, 0, s_2, m^2) &=& \frac{i}{4 \pi ^2} + \frac{s_2}{4 \pi ^4 \left(s-s_2\right)} \, D_2 + \frac{m^2}{2 \pi^4}\, \bar C_0 , \\
A_3 (s, 0, s_2, m^2) &=& - A_6 (s, 0, s_2, m^2) =  \nn \\
&& -\frac{i}{2 \pi ^2 \left(s-s_2\right)}
    -\frac{s_2}{2 \pi^4 \left(s-s_2\right)^2} \, D_2 -\frac{m^2}{\pi ^4 \left(s-s_2\right)} \, \bar C_0, \\
A_4 (s, 0,s_2, m^2) &=&\frac{1}{2 \pi ^4 \left(s-s_2\right)} \, D_2, \\
A_5 (s, 0,s_2, m^2) &=& - \frac{s_2} {\pi ^4 (s + s_2)^2}
\left(  s - 2 m^2  \right)  \, \bar C_0
- \frac{(s+s_2)}{2 \pi ^4  (s-s_2)^2}	\, \bar D_1 \nn \\
   && +\frac{(2  s+s_2) s_2}{\pi ^4 (s_2-s)^3} \, D_2-\frac{i s_2}{\pi ^2 (s-s_2)^2},
\eea
where $D_2$ is defined in Eq.(\ref{D_i}), while $\bar D_1$ and $\bar C_0 $ are the two $s_1 \rightarrow 0 $ limits of $D_1$ and $C_0 (s_1, s_2, s, m^2)$ respectively, that is
\bea
\bar D_1 &\equiv& \lim_{s_1\rightarrow 0} D_1 (s, s_1, m^2) =  i \pi^2 \left[ 2 - a_3 \log \frac{a_3 +1}{a_3 - 1}  \right], \\
\bar C_0 &\equiv& \lim_{s_1\rightarrow 0} C_0 (s, s_1, s_2, m^2) = -\frac{i \pi^2}{2(s-s_2)}
\left[ \log^2\frac{a_2 +1}{a_2 - 1}-\log^2\frac{a_3 +1}{a_3 - 1} \right].
\eea
The coefficients of the $w$'s in the L/T formulation, in this case, are
\bea
w_L (s, 0, s_2, m^2) &=& - \frac{4 i}{s} -\frac{8 m^2}{\pi ^2 s} \, \bar C_0,   \\
w_T^{(+)} (s, 0, s_2, m^2) &=& \frac{1 } {\pi ^2 (s- s_ 2)^2}
\biggl[   4 i \pi ^2 s
+ 2  (s + s_2 ) \,  \bar D_1
+  4 \, s \left ( 2 \, m^2 + s_2 \right)    \, \bar  C_0 \biggr.  \nn \\
&& \biggl.   \hspace{3.5cm} +\frac{2 \left(s^2+4 s_2  s + s_2^2\right) } {s -s_2} \, D_2 \biggr],   \\
w_T^{(-)} (s, 0, s_2, m^2) &=& - \frac{1 } {\pi ^2 (s- s_ 2)^2}
\biggl[    4 i \pi ^2 s
+ 2  (s + s_2 ) \, \bar  D_1
+ 4 \, s_2  \left ( 2 \, m^2 + s \right)  \, \bar C_0  \biggr.  \nn \\
&& \biggl. \hspace{3.5cm} +\frac{2 \left(s^2 - 6 s_2  s - s_2^2\right) } {s -s_2}\, D_2 \biggr],  \\
\tilde w_T^{(-)} (s, 0, s_2, m^2) &=& \frac{1 } {\pi ^2 (s- s_ 2)^2}
\biggl[ 4 i \pi ^2 s_2
+ 2  (s + s_2 ) \, \bar D_1
+  4 \, s_2  \left ( 2 \, m^2 + s \right) \, \bar C_0  \biggr.  \nn \\
&& \biggl. \hspace{3.5cm}  +\frac{2 \left(- s^2+6 s_2  s + s_2^2\right) } {s -s_2} \, D_2 \biggr].
\eea
Furthermore, in the case in which the massive amplitude has both external vector lines on-shell
\begin{itemize}
\item [e)] $k_1^2 =0$ \qquad $k_2^2 = 0$ \qquad $k^2 \neq 0$ \qquad $m\neq 0$
\end{itemize}
one obtains
\bea
A_1(0, 0, s, m^2)&=& -\frac{i}{4 \pi^2}\left( 1 + \frac{m^2}{s}\log^2\frac{a_3+1}{a_3-1}\right),\\
A_3(0, 0, s, m^2)&=& -A_6(0,0,s, m^2)= -\frac{i}{2 \pi^2 s}\left(1+ \frac{m^2}{s}\log^2\frac{a_3+1}{a_3-1}\right),\\
A_4(0, 0, s, m^2)&=& -\frac{i}{2 \pi^2 s}\left( a_3 \log\frac{a_3+1}{a_3-1} -2 \right).
\eea
These simple results are obtained with a limiting procedure, starting from the scalar triangle diagram with off-shell external lines and involves the function $\Phi(x,y)$ \cite{Usyukina:1992jd} already encountered in the explicit expression of the Rosenberg parameterization.
Instead, for the L/T parameterization we obtain
\bea
w_L(0,0,s,m^2)&=& -\frac{4 i}{s}\left[1+ \frac{m^2}{s}\log^2\left(\frac{a_3+1}{a_3-1}\right)\right],\\
w_T^{(+)}(0,0,s,m^2)&=& \frac{4 i}{s}\left[3+ \frac{m^2}{s}\log^2\left(\frac{a_3+1}{a_3-1}\right)-a_3 \log\left(\frac{a_3+1}{a_3-1}\right)\right],\\
w_T^{(-)}(0,0,s,m^2) &=& \tilde{w}_T^{(-)}(0,0,s,m^2) = 0.
\eea

Finally, the particles can be on-shell and both of mass $M$ and in this case we obtain 
\begin{itemize}
\item [f)] $k_1^2 =M^2$ \qquad $k_2^2 = M^2$ \qquad $k^2 \neq 0$ \qquad $m\neq 0$
\end{itemize}

\bea
A_1 (M^2, M^2, s, m^2 ) &=& -\frac{i}{4 \pi ^2} -\frac{ m^2}{2 \pi ^4} C_0,   \\
A_3 (M^2, M^2, s, m^2 ) &=& \frac{1}{\pi ^4 s \left(s-4 M^2\right)}
\left[   \frac{i \pi^2 } {2 } \left(2 M^2-s\right)  - \frac{ \left(2 M^2+s\right) M^2} {s-4 M^2 } D_M \right. \nn \\
&+& \left. \left( \frac{ 2 M^4 (M^2 - s )} {s-4 M^2 } - m^2 (s - 2 M^2 )    \right) C_0  \right], \\
A_4 (M^2, M^2, s, m^2 ) &=&
\frac{1}{\pi ^4 s \left(s-4 M^2\right)}
\left[   i \pi^2 M^2 +  \frac{ s^2 - 4 M^4} {2( s-4 M^2 )} D_M \right. \nn \\
&+& \left. \left (\frac{  M^2 (2 M^4 - 3 M^2 s + s^2 )} {s-4 M^2 } + 2 m^2 M^2 \right)   C_0  \right].
\eea
In the previous expressions we have denoted by $C_0$ the complete expression $C_0 (s_1, s_2, s, m^2)$ in Eq.(\ref{C0polylog}) computed at $s_1=s_2=M^2$. In addition to this we have defined 
\bea
D_M (M^2, s, m^2) &\equiv& B_0 (k^2, m^2) - B_0 (M^2, m^2) =
i \pi^2 \left[ a_M \log\frac{a_M +1}{a_M - 1} - a_3 \log \frac{a_3 +1}{a_3 - 1}  \right], \\
a_M &=& \sqrt{1 - \frac{4 m^2}{M^2}},  \qquad \qquad a_3 = \sqrt{1 - \frac{4 m^2}{s}}.
\eea
Similarly, the expressions of the $w$'s invariant amplitudes in the L/T parameterization for  the massive triangle amplitude are given by
\bea
w_L (s, m^2) &=& -\frac{4 i}{s} -\frac{8 m^2}{\pi ^2 s} C_0,  \\
w_T^{(+)} (s, m^2, M^2 ) &=& \frac{1 } {\pi ^2 (s- 4 M^2)}
\left[   4 i \pi ^2
+ \frac{4  (s + 2 M^2 )}{s - 4 M^2}  D_M
+  \left(8  m^2 + \frac{8 M^2 (s - M^2) }{s - 4 M^2 }\right ) C_0   \right], \nn \\ \\
w_T^{(-)} (s, m^2, M^2 ) &=& \tilde w_T^{(-)} (s, m^2, M^2 ) = 0.
\eea

\section{Appendix B. Results and conventions for the tensor reduction}
\label{PaVe}

We collect here some of the definitions and formulas that we have used in the main sections.
We have defined
\beq
B_0(k^2) = 
\int d^d q \frac{1}{(q-k)^2 \, q^2}
\label{B0k}
\eeq
and the unique scalar three-point function with all the momenta off-shell and $k$ ingoing, $k_1$,$k_2$ outgoing
\bea
C_0(k^2, k_1^2, k_2^2) &=&
\int d^d q  \frac{1}{(q-k)^2 \, (q-k_1)^2 \, q^2} = \frac{i \pi^2}{k^2} \Phi( x, y),
\label{C0}
\eea
where the  $\Phi(x,y)$ function is defined in Eq.~\ref{Phi}.
The explicit expression of the unrenormalized massless two-point scalar integrals in $d=4-2\eps$ with $\eps>0$  is

\beq
B_0 (k^2) =
 i \pi^2 \left[ \frac{1}{\bar \eps} + \log \bigg(\frac{ \mu^2}{k^2}\bigg) + 2    \right]
\label{twopoint}
\eeq
with a singular part in $ 1 / \bar \eps$, defined as
\beq
\frac{1}{\bar \eps} = \frac{1}{\eps} - \g - \ln \pi.
\eeq
The singularities in $1/\bar \eps$ and the dependence on the renormalization scale $\mu$ cancel out when taking into account the difference of two of these two-point scalar function $B_0$. 

The master integral used for the $m_i=m\neq0$ case is
\bea
C_0(k^2, k_1^2, k_2^2, m^2) &=&
\int d^d q  \frac{1}{((q-k)^2 -m^2)  \, ((q-k_1)^2-m^2) \, (q^2-m^2)} \nn \\
&=& - i \pi^2 \int_0^1 dw \int_0^w dz \frac{1}{b w^2 + a z^2 + c w z - b w - (a+c) z + m^2}
\eea
for the one-loop three-point function with $a = k_1^2$, $b = k_2^2$, $c = 2 k_1 \cdot k_2$. This parametric form of the scalar triangle has been used in the numerical check of our results for the form factors $A_i$ against those given by Rosenberg in \cite{Rosenberg:1962pp}.

The difference between two one-loop two-point functions has been defined in  Eq.(\ref{D_i}) as
\bea
D_i (s_i, s, m^2)&=& B_0 (k^2, m^2) - B_0 (k_i^2, m^2) = i \pi^2 \left[ a_i \log\frac{a_i +1}{a_i - 1} - a_3 \log \frac{a_3 +1}{a_3 - 1}  \right] \qquad i=1,2.
\nn \\
\eea
All the invariant amplitudes $A_i$ have been expressed as functions of $D_i$ introduced in Eq.(\ref{D_i}),
showing that the singularities coming from the two-point scalar functions and depending on the different momenta $k^2$, $k_1^2$ and $k_2^2$ perfectly cancel when inserted in the complete expansion of the invariant amplitudes $A_i$ for $\eps \rightarrow 0$. Notice that dimensional reduction and dimensional regularization with a partially anticommuting $\gamma_5$ give consistent answers for the anomaly loop with no need of a finite renormalization.

\end{document}